\documentclass[aip,twocolumn,letterpaper]{revtex4}

\pdfoutput=1  

\usepackage{fullpage}
\usepackage{xcolor}
\usepackage[normalem]{ulem} 

\usepackage{graphics}
\usepackage{epsfig}
\usepackage{url}

\begin{document}
\title{The rapid destruction of toroidal magnetic surfaces}
\author{Allen H Boozer}
\affiliation{Columbia University, New York, NY  10027\\ ahb17@columbia.edu}

\begin{abstract}

The operation of ITER will require reliable simulations in order to avoid major damage to the device from disruptions.  Disruptions are the sudden breakup of magnetic surfaces across the plasma volume---a fast magnetic reconnection.   This reconnection can be caused by the growth of perturbations outside of the plasma core causing an ideal perturbation to the core.  This causes an increasing ratio of the maximum to the minimum separation, $\Delta_{max}/\Delta_{min}$, between neighboring magnetic surfaces.  Magnetic reconnection becomes a dominant process when magnetic field lines can quickly interchange connections over a spatial scale $a_r$.  This occurs when $\Delta_{max}/\Delta_{min}\gtrsim a_r/\Delta_d$, where $\Delta_d$ is the scale over which non-ideal effects make magnetic field lines indistinguishable.  Traditional reconnection theory is fundamentally different.  It is a study of the steady-state cancellation of oppositely directed magnetic field components across a thin layer.  During more than sixty years, mathematical implications of Faraday's Law have been derived that clarify and constrain the physics of fast magnetic reconnection.  These are reviewed because they are not commonly known but are needed to understand and to place in context how an ideal magnetic evolution can cause reconnection to quickly become a dominant process no matter how small $\Delta_d/a_r$ may be. 

\end{abstract}

\date{\today} 
\maketitle

\begin{figure}
\centerline { \includegraphics[width=3.0in]{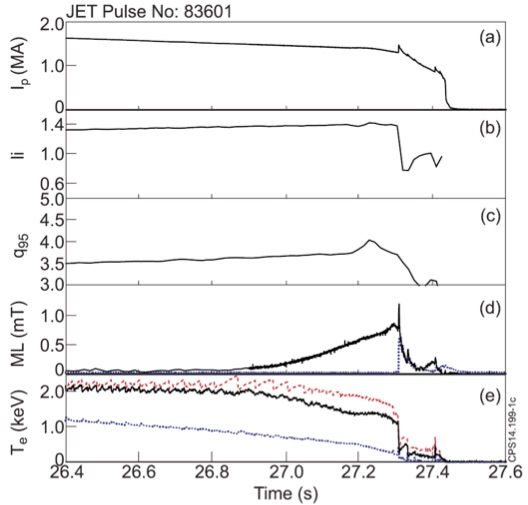} }
\caption{This is Figure 1 from \cite{de Vries:2016}, which gave the evolution of a JET plasma into a natural thermal quench.  Time traces are shown of (a) the  plasma current $I_p$, (b) the internal inductance $\ell_i$, (c) the safety factor near the plasma edge $q_{95}$, (d) the magnitude of magnetic perturbations, which may be due to tearing modes, and (e) the electron temperature at three radial positions in the plasma.  Reproduced from de Vries et al, Nucl. Fusion \textbf{56}, 026007 (2016), with the permission of the International Atomic Energy Agency.}  
\label{fig:JET}
\end{figure}

\section{Introduction}

Starting in 2029, the ITER tokamak is expected to operate at levels of plasma current at which disruptions could sufficiently damage the machine to require long shutdowns.   Proceeding empirically, by steering ITER plasmas away from such situations or by having a benign plasma shutdowns, will be difficult to impossible \cite{ITER-shutdown2018,Boozer:steering,Eidietis:2021}. The success of ITER will require that operational limits be defined computationally with extreme reliability.  Non-empirical methods of judging the reliability of simulation codes, such as those suggested in this paper, are needed for ITER to achieve its performance targets. 

During tokamak disruptions, magnetic surfaces are observed to be destroyed many orders of magnitude faster than would be naively expected from non-ideal effects.  Nonetheless,  magnetic surface breaking is impossible unless non-ideal effects are included in the equation for the evolution of magnetic fields.

The fast breakup of magnetic surfaces is an example of a fast magnetic reconnection.  Magnetic reconnection was defined in 1956 by Parker and Krook \cite{Parker-Krook:1956} as the ``\emph{severing and  reconnection of lines of force.}"  A magnetic reconnection  is called fast when it arises and progresses on a time scale primarily determined by an ideal-evolution timescale rather than the resistive timescale.  

Figure \ref{fig:JET} illustrates a naturally arising disruption on JET \cite{de Vries:2016}, which shows the current spike and the drop in the internal inductance, $\ell_i$.  These observations imply a large scale spreading of the current profile on a timescale $\sim1~$ms.  The rapid spread in the current profile requires an equally rapid rearrangement of the poloidal relative to the toroidal magnetic flux.

Naively, one would expect the time required for a major rearrangement of the fluxes in the plasma to be comparable $\tau_{L/R}$.  This is the time required for the poloidal flux in the plasma to be consumed by resistivity $\eta_0$ at the magnetic axis.   Appendix B of \cite{L/R} can be used to show that when the current profile is parabolic
\begin{equation}
\tau_{L/R} =\frac{3}{16\pi}\frac{2\kappa}{1+\kappa^2}\frac{\mu_0}{\eta_0}A,
\end{equation}
where $\kappa$ is the plasma elongation and $A$ is its cross sectional area.  

For the JET experiment illustrated  in Firgure \ref{fig:JET}, the central temperature was $\approx2$~keV and $A$ was $\approx 5$~m$^2$, so $\tau_{L/R} \approx 50$~s, which is $\approx5\times 10^4$ times longer than 1~ms.  The $\tau_{L/r}$ in ITER should be approximately ninety times longer, which is $\approx4.5\times 10^6$ times longer than 1~ms.

\begin{figure}
\centerline{ \includegraphics[width=2.5in]{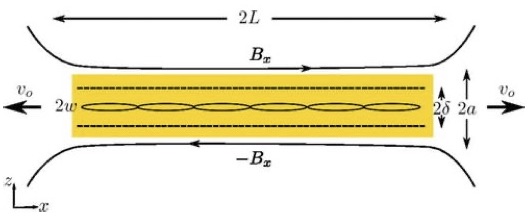}}
\caption{In plasmoid models, oppositely directed fields in the $B_x$ direction are pushed together forming a narrow current sheet.  Tearing instability of the sheet current creates the plasmoids, which are expelled at the Alfv\'en speed from the two ends of the current sheet. Reproduced from Y.-M. Huang, L. Comisso, and A. Bhattacharjee, Physics of Plasmas \textbf{26}, 092112 (2019), with the permission of AIP Publishing.}
\label{fig:plasmoid}
\end{figure}

As the ratio of the naively expected timescale divided by the actual timescale for flux rearrangement goes to infinity, what model should be used?  (1) The usual answer is the non-axisymmetric magnetohydrodynamics (MHD) codes, such as NIMROD, M3D-C1, and JOREK \cite{NIMROD,M3D-C1,JOREK}.  But, these have resolution limits that make realistic simulations of ITER difficult to impossible.  In any case, the physics must be understood to ensure the phenomena are being adequately represented.  (2)  The traditional model of magnetic reconnection is steady state with a sharp boundary between two regions in which a component of the magnetic field has the same magnitude but opposite signs.  The two-dimensional theory of ``\emph{General magnetic reconnection}"  was clarified by Schindler, Hesse, and Birn  \cite{Schindler:1988}.  The modern version is the plasmoid theory of reconnection \cite{plasmoid}, Figure \ref{fig:plasmoid}.  This model does not provide a good basis for understanding the fast breaking of magnetic surfaces during a tokamak disruption and has problematic mathematical issues, Appendix \ref{boundary-conditions}.  (3) A more useful model is a slowly growing non-axisymmetric but ideal magnetic perturbation that is applied to an initially axisymmetric tokamak.  What will be found is that when the perturbation grows beyond a critical amplitude, the preservation of magnetic surfaces becomes arbitrarily sensitive to non-ideal effects. 

Any magnetic evolution, $\partial \vec{B}/\partial t$, is determined by Faraday's law, which has subtle but important mathematical implications.  Many of these were published many years ago but are not commonly known in the plasma physics community.  A short summary of the ones needed to study toroidal magnetic surfaces are given in Section \ref{B-torus}.  A broader review that is needed to place this paper in context and to judge the credibility of different explanations of fast magnetic reconnection is given in Appendix \ref{sec:reconnection}.  

The concepts used in this paper are distinct from those used in traditional reconnection theory of a sharp change in magnetic field direction across a pre-formed current sheet, Figure \ref{fig:plasmoid}.  In tokamak disruptions and in many other reconnection problems, the initial state is smooth, but an ideal evolution carries the system into a state in which magnetic field lines change their connections over a large spatial scale $a_r$ despite non-ideal effects directly producing changes on only a small spatial scale, $\Delta_d$.   A recorded talk ``\emph{Fast Magnetic Reconnection}'' \cite{Boozer:JPP} my be helpful to those unfamiliar with this process or the mathematical results of Appendix \ref{sec:reconnection}.

Section \ref{sec:Ideal} briefly reviews what is known about the effect of growing ideal perturbations on magnetic surfaces.  A more extensive review is given in Appendix \ref{2D pert}.  This review places the notation that is used in the papers  \cite{RDR-kink,Hahm-Kulsrud,Boozer-Pomphrey,Zhou:2016,Zhou:2019,Huang:2021} in a toroidal context.  These papers stretch back over several decades.  As discussed in recent papers  \cite{Zhou:2019,Huang:2021}, even two-dimensional, time-independent models are difficult numerically.  

What is needed for magnetic-surface breaking during tokamak disruptions is a far more challenging situation in which  the solutions for ideal perturbations are three dimensional and time dependent. Fine scale structures are spread over the entire plasma rather than limited to a narrow region near a single rational surface as in two-dimensional models.  It is not known how large a ratio of timescales the three-dimensional MHD codes can credibly represent.  These codes have not been used to study growing ideal perturbations to magnetic surfaces in either two or three dimensions, which would clarify their practical limitations.

Section \ref{sec:B-torus} sketches the theory of magnetic fields in a torus and explains how the trajectory of a single magnetic field line that lies in an irrational magnetic surfaces gives the ratio $\Delta_{max}/\Delta_{min}$ of the greatest to the least separation between the neighboring magnetic surfaces.  A fast breakup or reconnection of magnetic surfaces arises when $\Delta_{min}$ equals the distinguishability distance $\Delta_d$ with $\Delta_{max}$ equal to the radial scale $a_r$ over which magnetic reconnection causes a major change in the plasma equilibrium.

Section \ref{sec:distinguishability} explains the two distances $\Delta_d$ and $a_r$ that in effect define magnetic reconnection.   $\Delta_d/a_r$ is extremely small in many cases of magnetic fields embedded in natural or laboratory plasmas.  Nevertheless, whenever an ideal evolution causes $\Delta_{max}/\Delta_{min}$ to become larger than $a_r/\Delta_d$, magnetic reconnection becomes a dominant process.

Section \ref{Implications} summarizes and gives the importance of the new results of the paper.  It also discusses simulations that should be carried out based on the methods developed in this paper and the relation of this paper to other theories of magnetic reconnection.

Two topics that are important for understanding this paper have been discussed in a number of papers in prominent journals over many years.  Nonetheless, knowledge of these topics  is not widespread in the plasma physics community.  The two appendices provide the required background information. 

Appendix \ref{sec:reconnection} reviews the mathematical implications of Faraday's law that have been developed over more than sixty years.  These are constraints on all magnetic evolutions, including magnetic reconnection.

Appendix \ref{2D pert} reviews the literature on the effects on magnetic surfaces of a growing ideal magnetic perturbation beyond the brief summary given in Section \ref{sec:Ideal}.  In particular, it explains the relation between toroidal systems and the simple two-dimensional slab models used in the literature.


\section{Ideal perturbations to tokamaks \label{sec:Ideal}} 

As illustrated in Figure \ref{fig:JET} of an experiment in JET, MHD perturbations grow over hundreds of milliseconds, but the evolution suddenly changes its timescale from hundreds to approximately one millisecond. Changes in the large-scale properties, the internal inductance $\ell_i$ and the current spike, primarily occur on the one millisecond timescale.  This experiment can have a number of interpretations.  Nevertheless, since hundreds of milliseconds is fast compared to the natural scale for flux rearrangements in the central part of the plasma, it is reasonable to ask what the response of the central magnetic surfaces would be to a growing ideal perturbation.

The extensive literature on ideal perturbations \cite{RDR-kink,Hahm-Kulsrud,Boozer-Pomphrey,Zhou:2016,Zhou:2019,Huang:2021} is unfamiliar to many in plasma physics.  Appendix \ref{2D pert} gives more details for those who find this section too brief.

Since toroidal magnetic surfaces cannot be broken by an ideal perturbation \cite{Boozer:RMP}, the strength of the perturbation can be measured by how far $\xi(\psi,\theta,\varphi,t)$ the surfaces are displaced; $\psi$ is the toroidal magnetic flux enclosed by a magnetic surface, $\theta$ is a poloidal, and $\varphi$ is a toroidal angle.  The displacement of a surface means along the normal to the surface, the direction $\hat{n}\equiv\vec{\nabla}\psi/|\vec{\nabla}\psi|$.  The velocity with which the magnetic surface is displaced is $\vec{u}_s=\hat{n}\partial \xi/\partial t$.   The velocity of the magnetic field lines is $\vec{u}_\bot=\vec{u}_s+\vec{u}_t$, where the tangential field line velocity $\vec{u}_t$ is in the magnetic surface, in the $\hat{b}\times\hat{n}$ direction where $\hat{b}\equiv \vec{B}/B$.

The subtlety of ideal displacements is their behavior near rational surfaces, surfaces $\psi=\psi_s$ on which the rotational transform is the ratio of two integers $\iota(\psi_s)=M_s/N_s$; the field lines close on themselves after $M_s$ toroidal and $N_s$ polodial circuits of the torus.  Shielding currents must arise to prevent the resonant part of the perturbation from splitting a rational surface to form an island.  The resonant part has an angular dependence  $\cos\big(j_h(M_s\theta-N_s\varphi)\big)$ or the sinusoidal equivalent with $j_h$ an integer.

When a displacement is suddenly initiated, the amplitude of the shielding current starts  as essentially spatially uniform within the region enclosed by magnetic surfaces that are too close to the rational surface for a shear Alfv\'en wave to have had time to cover the surface \cite{Hahm-Kulsrud}, Equation (\ref{j-density}).  The width of this region scales inversely with time, $\sim r_s\tau_A/t$, where $r_s$ is the radius of the rational surface and $\tau_A$ is an Alfv\'en transit time, Equation (\ref{current width}).  The plasma response is linear, during the time that the resonant part of $\xi$ is smaller than $\sim r_s\tau_A/t$, and this short-time response was the only one correctly calculated by Hahm and Kulsrud in \cite{Hahm-Kulsrud}.

\begin{figure}
\centerline { \includegraphics[width=3.0in]{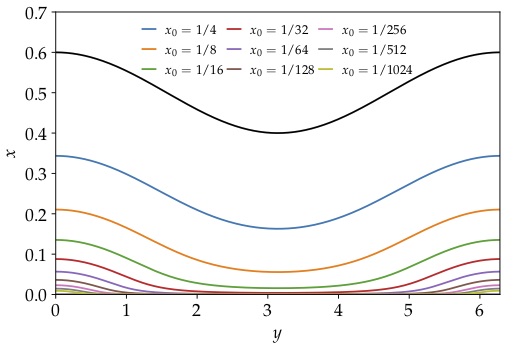} }
\caption{This figure gives the perturbed shape of magnetic surfaces that had different initial separations $x_0$ from the rational surface $\iota(\psi_s)=M_s/N_s$ in response to a perturbation $\cos(M_s\theta-N_s\varphi)$.  The distance of perturbed  surfaces from the rational surface, $x$, is normalized by the amplitude of the perturbation just outside the current channel, and  $y\equiv M_s\theta-N_s\varphi$.  Reproduced from Y.-M. Huang, S. R. Hudson, J. Loizu, Y. Zhou, and A. Bhattacharjee,  arXiv (August 2021),  $\big<$https://arxiv.org/pdf/2108.09327.pdf$\big>$, with permission.  }  
\label{fig:Displacement}
\end{figure}

The plasma response becomes much more complicated once the current channel width is comparable to $\xi$.  The non-linear response is sufficiently complicated that calculations have been limited to helically symmetric systems in the $t\rightarrow\infty$ limit of the plasma response to a perturbation that is held at a constant amplitude, \cite{Boozer-Pomphrey,Zhou:2016,Zhou:2019,Huang:2021}.  The analyses are based  on a method originally developed for the $M=1,N=1$ ideal kink by Rosenbluth, Dagazian, and Rutherford \cite{RDR-kink}.  In these studies, the parallel current density $j_{||}$ becomes singular at the resonant rational surface, but approximately half of the shielding current is spread over a channel with a width comparable to the amplitude of the displacement just outside the current channel.   The deformation of surfaces by the perturbation is illustrated in Figure \ref{fig:Displacement}.  A magnetic surface that had an initial spatial separation $x_0$ from the rational surface will have separation that varies from $\Delta_{max}$ to $\Delta_{min}$ with ratio of separations becoming infinite as the rational surface is approached, $\Delta_{max}/\Delta_{min}\propto 1/x_0^{4/3}$.  The closest approach $\Delta_{min}$ scales $x_0^2$ and a farthest $\Delta_{max}$ scales as $x_0^{2/3}$ in the region in which $x_0$ is small compared to the imposed displacement.  The width of the region of farthest separation scales as $x_0^{1/3}$, so the cross-sectional area of this region scales as $x_0$.  When the Fourier decomposition of $\xi$ has a single harmonic, $j_h=1$, just outside the current channel, the Fourier decomposition of the displacement has increasing number of harmonics the closer the unperturbed surface was to the rational surface.

The short-time linear studies and force-balance considerations imply the current density is finite for finite time, apparently the maximum current density increases linearly with time, Appendix \ref{2D pert}. 

The nature of the displacement $\xi(\psi,\theta,\varphi,t)$ becomes far more complicated in a toroidal tokamak than in the helically symmetric systems that have been studied.  Toroidicity and the axisymmetric shaping drive  poloidal or $M$-number harmonics.  The production of the $j_h$ harmonics by the interaction with a single rational surface and the spread in the $M$ harmonics by toroidicity imply that even an external perturbation that is a pure Fourier mode, which resonates with a single magnetic surface, produces a broad spectrum of helicities once it interacts with the plasma.  Calculations that include many helicities and consequently many resonant rational surfaces are far more difficult than single-helicity calculations.  

What is clear is that an externally driven displacement $\xi_d$ that has an increasing amplitude and a resonate interaction with at least one rational surface will over time involve an ever greater number of rational surfaces in the plasma response.  The ratio between the maximum and minimum separation between neighboring surfaces becomes large as $\xi_d$  increases.  This effect becomes extreme when the displacement $\xi_d$ is comparable to the distance between low order rational surfaces, which gives a heuristic condition for a fast magnetic reconnection to occur.  The implication is that a fast reconnection in a tokamak is similar to a fast reconnection sparked by the motion of magnetic footpoints \cite{Rec-example} and Appendix \ref{sec:reconnection}.  An ideal evolution can not break magnetic field lines but can make their large-scale breaking result from arbitrarily small non-ideal effects.


\section{Magnetic fields in a torus \label{sec:B-torus} }

 
 \subsection{Mathematical properties of $\vec{B}$ \label{B-torus}}
 
  The mathematical properties of magnetic fields in a torus were reviewed in 2004 in Section III of Reference \cite{Boozer:RMP}.   This section is a short summary.  A broader perspective is given in Appendix \ref{sec:reconnection}.
 
A time dependent magnetic field can be written in terms of the toroidal $\psi$ and the poloidal $\psi_p(\psi,\theta,\varphi,t)$ magnetic fluxes and the position vector $\vec{x}(\psi,\theta,\varphi,t)$.  The angle $\theta$ is a poloidal, the short way around the torus, and $\varphi$ is a toroidal, long way around the torus, angle. 
 \begin{eqnarray}
2\pi\vec{B}&=&\vec{\nabla}\psi \times\vec{\nabla}\theta + \vec{\nabla}\varphi\times\vec{\nabla}\psi_p(\psi,\theta,\varphi,t) \label{B-gen}
\end{eqnarray}
The poloidal flux is the Hamiltonian that gives the magnetic field lines in its canonical coordinates, which are $(\psi,\theta,\varphi)$, at fixed points in time:
\begin{equation}
\frac{d\psi}{d\varphi}=- \frac{\partial\psi_p}{\partial\theta} \hspace{0.2in}\mbox{and}\hspace{0.2in} \frac{d\theta}{d\varphi}=+ \frac{\partial\psi_p}{\partial\psi}.
\end{equation}
A change in magnetic field line topology requires a change in $\psi_p$.  The canonical coordinates can always be chosen as time advances so
 \begin{eqnarray}
 &&\frac{\partial  \psi_p(\psi,\theta,\varphi,t)}{\partial t}=V,  \hspace{0.2in}\mbox{where} \\
 && V\equiv \lim_{L\rightarrow\infty} \frac{\int^L_{-L} \vec{E}\cdot d\vec{\ell}}{\int^L_{-L} \vec{\nabla}\left(\frac{\varphi}{2\pi}\right)\cdot d\vec{\ell}}
 \end{eqnarray}
is the loop voltage and $d\vec{\ell}$ is the differential distance along $\vec{B}$.  The generalization of the magnetic field line velocity is $\vec{u}_c=\partial\vec{x}(\psi,\theta,\varphi,t)/\partial t$, which is the velocity of the canonical coordinates  through space.  

A magnetic evolution is ideal when and only when the loop voltage $V$ is zero.
 
 A number of corollaries follow \cite{Boozer:RMP}: (1) The tying together of the plasma and the magnetic field in an ideal evolution is actually two results. The stronger is the tying together of the toroidal and the poloidal magnetic flux, which occurs when the loop voltage $V=0$.  The weaker is the tying of the plasma to either the toroidal or the poloidal flux.  (2) Magnetic surfaces are perfect when $\psi_p$ is a function of only $\psi$ and time.  (3) Magnetic surfaces can be broken only on rational surfaces, which means places where magnetic field lines are closed---bite their own tail.  On irrational surfaces, each field line comes arbitrarily close to every point on the surface, so $V$ is a function of only $\psi$ and $t$. (4) The velocity of the plasma $\vec{v}$ has no direct role in changes in the magnetic topology of magnetic field lines and, therefore, in magnetic reconnection.

This paper explains how an ideal magnetic evolution, $V=0$, can enhance the sensitivity of magnetic-surface preservation to non-ideal effects by an arbitrarily large number of orders of magnitude.   This follows from the properties of an externally applied ideal perturbation that has a steadily increasing magnitude to a tokamak with perfect magnetic surfaces.  The hypersensitivity to non-ideal effects comes through the velocity of the canonical coordinates, $\vec{u}_c$ since $\psi_p(\psi,t)$ is time independent in an ideal evolution.


\subsection{Representation of a $\vec{B}$ with surfaces}

A magnetic field has surfaces when a function $\psi(\vec{x},t)$ exists, with $\vec{\nabla}\psi\neq0$ except at isolated points, such that $\vec{B}\cdot\vec{\nabla}\psi=0$. 

Wherever magnetic field lines lie on nested surfaces that are spatially bounded, these surfaces must be toroidal and can be denoted by the toroidal magnetic flux $\psi$ that each surface encloses.  Equation (\ref{B-gen}) can then be written in the magnetic coordinate form \cite{Boozer:RMP}
\begin{eqnarray}
2\pi\vec{B}&=&\vec{\nabla}\psi \times\vec{\nabla}\theta + \iota(\psi)\vec{\nabla}\varphi\times\vec{\nabla}\psi.\label{B-mag}
\end{eqnarray}
Coordinates $(\psi,\theta,\varphi)$ that are consistent with this equation are called magnetic coordinates.

 The toroidal angle $\varphi$ can be taken to be the polar angle of $(R,\varphi,Z)$ cylindrical coordinates.   The equation for a magnetic field line that was at $(\psi_0,\theta_0)$ at $\varphi=0$ is $\psi=\psi_0$ and $\theta=\theta_0+\iota(\psi_0)\varphi.$   The rotational transform $\iota=(\partial\psi_p/\partial\psi)_t$ is the inverse of the safety factor, $q=1/\iota$.  The magnetic field can also be written in what is called the Clebsch form, $2\pi\vec{B}=\vec{\nabla}\psi \times\vec{\nabla}\theta_0$.

The position in space associated with a given point in magnetic coordinates, $\vec{x}(\psi,\theta,\varphi)$ can be given using cylindrical coordinates $(R,\varphi,Z)$ with 
\begin{equation}
\vec{x}(\psi,\theta,\varphi) = R(\psi,\theta,\varphi)\hat{R}(\varphi) + Z(\psi,\theta,\varphi)\hat{Z}. \label{cyl}
\end{equation}
The three tangent vectors of magnetic coordinates are
\begin{eqnarray}
\frac{\partial \vec{x}}{\partial \psi} &=& \frac{\partial R}{\partial\psi}\hat{R} + \frac{\partial Z}{\partial \psi}\hat{Z}, \\
\frac{\partial \vec{x}}{\partial \theta} &=& \frac{\partial R}{\partial\theta}\hat{R} + \frac{\partial Z}{\partial \theta}\hat{Z}, \mbox{   and  } \\
\frac{\partial \vec{x}}{\partial \varphi} &=& \frac{\partial R}{\partial\varphi}\hat{R} + \frac{\partial Z}{\partial \varphi}\hat{Z} + R\hat{\varphi},
\end{eqnarray}
where the three orthonormal unit vectors of cylindrical coordinates satisfy $\hat{Z}\times\hat{R} =\hat{\varphi}$ and $d\hat{R}/d\varphi=\hat{\varphi}$.


\subsection{Separation between magnetic surfaces \label{Sec:Sep}}

This section will derive relations, which are required to prove Equation (\ref{reciprocal}).  Equation (\ref{reciprocal}) relates the separation of neighboring magnetic surfaces in constant-$\varphi$ planes to the distance between a pair of neighboring field lines that lie on the same magnetic surface. The variation in the separation of a pair of lines in a magnetic surface is more easily ascertained than is the variation in the separation between neighboring magnetic surfaces.

Calculating the separation of neighboring magnetic surfaces in constant-$\varphi$ planes is a major simplification to the calculation of the actual separation of neighboring surfaces.  Nevertheless, it is an excellent approximation because the strength of the toroidal field $B_\varphi$ in tokamaks, $|\vec{B}-B_\varphi\hat{\varphi}|/B_\varphi \lesssim0.1$, makes it difficult to produce sharp toroidal bends in the surfaces.

Equation (\ref{B-mag}), which gives the magnetic field in magnetic coordinates, and the expressions for tangent vectors imply
\begin{eqnarray}
&&2\pi\vec{B}\cdot\frac{\partial\vec{x}}{\partial\psi}\times \frac{\partial\vec{x}}{\partial\theta} = 1. \label{B-dot}
\end{eqnarray}
The derivation uses the orthogonality relations of general coordinates, which are derived in the Appendix of \cite{Boozer:RMP}: $\vec{\nabla}\psi \cdot \partial\vec{x}/\partial\psi=1$, $\vec{\nabla}\theta \cdot \partial\vec{x}/\partial\psi=0$, etc.  Equation (\ref{cyl}), which gives the positions associated with points in magnetic coordinates, and Equation (\ref{B-dot}) imply
\begin{eqnarray}
&&\frac{\partial\vec{x}}{\partial\psi}\times \frac{\partial\vec{x}}{\partial\theta} = \frac{\hat{\varphi}}{2\pi B_\varphi},  \mbox{   or  }  \\
&&\Big|\frac{\partial\vec{x}}{\partial\psi}\Big| \Big|\frac{\partial\vec{x}}{\partial\theta}\Big| \Big|\sin\alpha\Big| = \frac{1}{2\pi B_\varphi}
\end{eqnarray}
where $B_\varphi\equiv \vec{B}\cdot\hat{\varphi}$ is the toroidal magnetic field and $\alpha$ is the angle between the tangent vectors $\partial\vec{x}/\partial\psi$ and $\partial\vec{x}/\partial\theta$.  This angle is defined by 
\begin{eqnarray}
\cos\alpha\equiv\frac{ \frac{\partial\vec{x}}{\partial\psi} \cdot \frac{\partial\vec{x}}{\partial\theta} }{
 \left|\frac{\partial\vec{x}}{\partial\psi} \right| \left| \frac{\partial\vec{x}}{\partial\theta}\right| }. \label{alpha}
\end{eqnarray} 

The separation $\Delta_\psi \delta\psi$ between a pair of neighboring magnetic surfaces, one at $\psi$ and the other at $\psi+\delta\psi$ in a constant-$\varphi$ plane is calculated using
\begin{equation}
\vec{\Delta}_\psi \delta\psi=  \frac{\partial\vec{x}}{\partial\psi} \delta\psi + \frac{\partial\vec{x}}{\partial\theta} \delta\theta
\end{equation} 
with $\delta\theta/\delta\psi$ chosen to minimize the separation $\Delta_\psi\equiv\sqrt{\vec{\Delta}_\psi \cdot \vec{\Delta}_\psi}$.  Since
\begin{eqnarray}
&&\Delta_\psi^2 = \left(\frac{\partial\vec{x}}{\partial\psi}\right)^2 + 2 \frac{\delta\theta}{\delta\psi}\frac{\partial\vec{x}}{\partial\psi}\cdot\frac{\partial\vec{x}}{\partial\theta}+ \left(\frac{\delta\theta}{\delta\psi}\right)^2\left(\frac{\partial\vec{x}}{\partial\theta}\right)^2, \nonumber \\ \\
&&\frac{\delta\theta}{\delta\psi}= - \frac{\left| \frac{\partial\vec{x}}{\partial\psi} \right|}{\left| \frac{\partial\vec{x}}{\partial\theta} \right|} \cos\alpha  \label{ratio at min}
\end{eqnarray}
at the minimum of $\Delta_\psi^2$ using Equation (\ref{alpha}) for $(\partial \vec{x}/\partial\psi)\cdot(\partial \vec{x}/\partial\theta)$.  Consequently,
\begin{eqnarray}
&&\Delta_\psi   =    \left| \frac{\partial\vec{x}}{\partial\psi}\right| |\sin\alpha|  \hspace{0.2in} \mbox{     and   }\label{Delta-psi} \\
&&\Delta_\psi \Big|\frac{\partial\vec{x}}{\partial\theta}\Big| = \frac{1}{2\pi B_\varphi}. \label{Delta-psi}
\end{eqnarray}
The term ``neighboring surfaces" implies the limit $\delta\psi\rightarrow0$.  Although the separation between neighboring surfaces $\Delta_\psi$ is of primary interest for magnetic reconnection, the spatial separation between a pair of magnetic field lines that are both in the same magnetic surface is far easier to determine given a magnetic field $\vec{B}(\vec{x})$.


\subsection{Field-line separation}

A single field-line integration, $d\vec{x}/d\varphi =  \vec{B}(\vec{x})/\vec{B}\cdot\vec{\nabla}\varphi$, where $\vec{\nabla}\varphi=\hat{\varphi}/R$, allows one to determine the separation between a pair of lines in a magnetic surface that have an initial separation $\delta(0)$.  $\vec{B}(\vec{x})$ is assumed to be given in $(R,\varphi,Z)$ cylindrical coordinates.  First, integrate the field line equation starting at $\varphi=0$ and determine $\vec{x}(\varphi)$.  Since almost all surfaces are irrational, $\vec{x}(\varphi=2\pi k)$, with $k$ an integer, will come arbitrarily close to $\vec{x}(\varphi=0)$ for some value $k=k_0$.  Let 
\begin{eqnarray}
\delta(\varphi) &\equiv& |\vec{x}(\varphi+2\pi k_0)-\vec{x}(\varphi)| \label{delta-varphi}
\end{eqnarray} 
with $\delta(0)\equiv\delta(\varphi=0)$ sufficiently small that the ratio $\delta(\varphi)/\delta(0)$ achieves its asymptotic limit. Define
\begin{eqnarray}
\frac{\Delta_\theta(\varphi)}{\Delta_\theta(0)}  \equiv  \frac{B_\varphi(\varphi) \delta(\varphi)}{B_\varphi(0)\delta(0)}. \label{Delta-theta}
\end{eqnarray}

Although there is only one field-line integration, a pair of neighboring field lines are being followed:  one at $\theta_1=\theta_0 + \iota\varphi$ and the other at $\theta_2=\theta_0 + \delta\theta_0 + \iota\varphi$. The separation $\delta(0)$ is proportional difference in the  poloidal angle $\delta\theta_0$ at $\varphi=0$;
\begin{equation}
\delta(0) = \Big|\frac{\partial\vec{x}}{\partial\theta}\Big|_{\theta_0}\delta\theta_0.
\end{equation}
The difference $\delta\theta_0$ is unknown, but that is irrelevant when $\delta\theta_0$ sufficiently small compared to the $\theta$ range through which a significant change in $\vec{x}(\psi,\theta,\varphi)$ occurs.  

Since $\iota(\psi)$ is irrational on almost all magnetic surfaces, a single field line $\theta=\theta_0 + \iota\varphi$, will come arbitrary close to every point on the surface, and the ratio of Equation (\ref{Delta-theta}), $\delta(\varphi)/\delta(0)$, becomes independent of $\delta\theta_0$ for $\delta(0)$ sufficiently small.

Equation (\ref{Delta-theta}) is equivalent to
\begin{eqnarray}
\frac{\Delta_\theta(\varphi)}{\Delta_\theta(0)} =  \frac{B_\varphi(\varphi) \Big|\frac{\partial \vec{x}}{\partial\theta} \Big| }{B_\varphi(0) \Big|\frac{\partial \vec{x}}{\partial\theta} \Big|_{\theta=\theta_0} }.  \label{Delta-theta-calc}
\end{eqnarray}
Equation (\ref{Delta-psi}) implies
\begin{eqnarray}
\frac{\Delta_\psi(\varphi)}{\Delta_\psi(0)}=\frac{\Delta_\theta(0)}{\Delta_\theta(\varphi)},  \label{reciprocal}
\end{eqnarray}
where $\Delta_\theta(\varphi)$ can be determined from a single magnetic-field line, $\vec{x}(\varphi)$ using Equations (\ref{delta-varphi}) and (\ref{Delta-theta}).

Where two neighboring magnetic field lines are close to each other on a magnetic surface, which means $\Delta_\theta$ small, the magnetic surfaces are far part, which means $\Delta_\psi$ is large, and vice versa.  This phenomenon is well known as the separatrix that defines a tokamak divertor is approached from the plasma side.


\subsection{Variation in the separation of surfaces \label{surf-sep-var}}

When the ratio of the maximum to the minimum separation of neighboring magnetic surfaces, $\Delta_{max}/\Delta_{min}$ becomes sufficiently large, $\Delta_{max}/\Delta_{min}\gtrsim a_r/\Delta_d$, reconnection becomes inevitable.  

To determine $\Delta_{max}/\Delta_{min}$, define the ratio function
\begin{eqnarray}
\tilde{r}(\varphi) &\equiv& \frac{\Delta_\theta(\varphi)}{\Delta_\theta(0)}, \mbox{  or equivalently  } \\
&=&  \frac{\Delta_\psi(0)}{\Delta_\psi(\varphi)}
\end{eqnarray}
using Equation (\ref{reciprocal}).  

The variation in the separation is given by
\begin{equation}
\frac{\Delta_{max}}{\Delta_{min}} \equiv \frac{\tilde{r}_{max}}{\tilde{r}_{min}}, \label{tilde-r}
\end{equation}
where $\tilde{r}_{max}$ and $\tilde{r}_{min}$ are the maximum and the minimum value of $\tilde{r}(\varphi)$ on a magnetic surface.

The calculation of $\Delta_{max}/\Delta_{min}$ requires following an arbitrarily chosen magnetic field line on a $\psi$-surface for a sufficient number of toroidal transits for the line to come close to every point on that surface---in particular, it must come close to the points where $\tilde{r}_{max}$  and $\tilde{r}_{min}$ are located.

Reconnection only requires non-ideal effects destroy the distinguishability of magnetic field lines where they are closer together than the minimum distances $\Delta_d$, over which they are distinguishable, Section \ref{sec:distinguishability}.  Consequently, the distinguishability of neighboring magnetic surfaces is also destroyed if they come closer to each other than $\Delta_d$ at any point in space.  Any pair of neighboring lines in a magnetic surface  have their maximum separation at the point at which its neighboring magnetic surfaces have their closest approach.


\section{The  reconnection distances $\Delta_d$ and $a_r$ \label{sec:distinguishability}} 

Two distances define when magnetic reconnection is important.  The distinguishability distance, $\Delta_d$, is the minimum separation two magnetic field lines can have before they become indistinguishable.   An ideal evolution has $\Delta_d=0$ by definition, but when magnetic fields are embedded in real plasmas, non-ideal effects, such as resistivity and electron inertia, always make $\Delta_d$ non-zero.  The critical scale for reconnection, $a_r$, is the minimum distance required for the breaking of field line connections to change the speed or nature of the magnetic evolution. 

Naively one would assume that an evolution would be approximately ideal when $\Delta_d/a_r<<1$, but this is false.  The reason is that an ideal evolution can cause pairs of field lines to have a ratio of maximum to minimum separation, $\Delta_{max}/\Delta_{min}$ that increases exponentially with time, so $\ln(\Delta_{max}/\Delta_{min}) \propto t/\tau_{ev}$, where $\tau_{ev}$ is the ideal evolution timescale.  When $\Delta_{max}/\Delta_{min}\gtrsim a_r/\Delta_d$, pairs of field lines can interchange connections on the scale $a_r = \Delta_{max}$ by losing their distinguishability at their location of closest approach $\Delta_{min}\lesssim\Delta_d$.  The required time for reconnection to occur is $\approx \tau_{ev}\ln(a_r/\Delta_d)$.  Even extreme assumptions about the smallness of $\Delta_d/a_r$ are consistent with reconnection becoming a dominant effect in approximately ten ideal evolution times, $\tau_{ev}$.  Appendix \ref{sec:reconnection} reviews the mathematics, \cite{Rec-example} gives a numerical example, and \cite{Boozer:rec-phys} relates this reconnection phenomenon to the equilibration of temperature in a room in approximately ten minutes instead of the couple weeks that would be naively expected from thermal diffusion. 

The naive assumption that  $a_r/\Delta_d>>1$ would give an approximately ideal evolution is certainly not what is observed in nature and in the laboratory.   Examples are from the solar corona \cite{Amitava:2004} to tokamak disruptions \cite{de Vries:2016}.  Magnetic field lines do not change connections freely.  Nevertheless, even when $a_r$ is many orders of magnitude greater than $\Delta_d$, the time required for reconnection to fundamentally change the evolution is often only an order of magnitude longer than the time scale of the ideal magnetic evolution itself.  The best known example is Parker's observation \cite{Parker:1973} that reconnection is commonly observed to proceed at the speed $\approx0.1V_A$, where the Alfv\'en speed is the natural speed of an ideal evolution when static force balance is lost.

Magnetic field lines are determined at fixed points in time.  In an ideal magnetic evolution, the magnetic field lines at one point in time can be mapped into the lines at another.  In a torus with magnetic surfaces, the time dependent position vector $\vec{x}(\psi,\theta,\varphi,t)$ together with the time independent poloidal flux, $\psi_p(\psi)$, provide the mapping. Non-ideality and magnetic reconnection are produced to the extent that this mapping is broken. 

Numerical as well as physical effects make $\Delta_d$ non-zero.  Pariat and Antiochos  \cite{Pariat-Antiochos} explicitly used the finite grid scale to obtain reconnection in a simulation.  Figure 3 of Huang et al \cite{Huang:2021} illustrates a study of an ideal magnetic perturbation in helically symmetry.  Their figure shows that an inadequate resolution in the radial direction coupled with a Fourier decomposition in the periodic direction can produce magnetic surface overlap.  Surface overlap implies reconnection.  However, avoiding surface overlap by constraining the contortions of the magnetic surfaces tends to artificially enhance their robustness to small surface-breaking effects.     
 
Since the distinguishability distance is defined by an inequality, expressions for $\Delta_d$ are valid if they satisfy this inequality.  When $\Delta_d$ is small and enters logarithmically, as is it generally does in the mathematics of a near-ideal evolution, Appendix \ref{sec:reconnection}, the precise numerical value for $\Delta_d$ makes little difference in the predicted time for magnetic reconnection to ensue.  
 
 Electron inertia gives a distinguishability distance equal to the electron skin depth, $c/\omega_{pe}$, which is similar in effect to that of a finite spatial grid.  A detailed derivation is given in Appendix C of \cite{Boozer:null-X}.
 
 Plasma resistivity $\eta$ diffuses magnetic field lines with a diffusion coefficient $\eta/\mu_0$.   The expected distinguishability distance from this effect is $\Delta_d^2 =(\eta/\mu_0) t$.  The shortest time that can characterize lines not interdiffusing is $\Delta_d/u_\bot$. Consequently, if two magnetic field lines come closer than $\Delta_d=\eta/\mu_0u_\bot$ at any point on their trajectories, the two lines must be indistinguishable.  
 
When combined with the references \cite{Rec-example,Boozer:rec-phys}, Equation (\ref{B-ev}) for the evolution of a magnetic field gives a different way to estimate for the resistive distinguishability distance, $\Delta_d=\eta/\mu_0u_\bot$.  This argument is simple when terms that are logarithmic in $\ln(a_u/\Delta_d)$ are taken to be of order unity, where $a_u$ is the spatial scale over which the field-line velocity changes, which is assumed to satisfy $a_r\lesssim a_u$.   The ideal evolution contributes to $\partial\vec{B}/\partial t$ as $|\vec{\nabla}\times(\vec{u}_\bot\times\vec{B})|\sim u_\bot B/a_u$.  The resistivity contributes to $\mathcal{E}$ as $\eta\left<j_{||}\right>$, with $\left<j_{||}\right>$ the average of the parallel current density along the line,  and contributes to $\partial\vec{B}/\partial t$ as $\eta |\vec{\nabla}\left<j_{||}\right>|\sim \eta \left<j_{||}\right>/\Delta_d$.  The current flows in ribbons along the magnetic field lines that become exponentially narrower in one direction across the magnetic field lines \cite{Rec-example,Boozer:rec-phys}, Appendix \ref{sec:gen-ev}, until the ribbons become so thin that the effect of resistivity on the evolution is as fast as that of the ideal evolution.  The current ribbons are exponentially broader in the other direction across the lines with $\left<j_{||}\right>\sim B/\mu_0a_r$ evolving only logarithmically, by an amount proportional to $\sim\ln(a_r/\Delta_d)$ before reconnection becomes a dominant process.

The distinguishability distance $\Delta_d$ is a combination of the electron inertia and resistive effects.  Since the electron skin depth is $c/\omega_{pe} = 5.31 \times 10^{-4} \sqrt{10^{20}~\mbox{m}^{-3}/n}$~m, and the magnetic diffusivity $\eta/\mu_0 = 6.6 \times 10^{-4}~\mbox{m$^2$/s} ~(10~\mbox{keV}/T)^{3/2}$,  the qualitative form of the distinguishability distance is
\begin{eqnarray}
\Delta_d &\approx& \frac{c}{\omega_{pe}} \Big( 1 + \frac{u_c}{u_\bot}\Big) \hspace{0.3in}  \mbox{   with    } \label{comb-Delta_d} \\
u_c &\equiv& \frac{\eta/\mu_0}{c/\omega_{pe}} \\
&=&1.2~\mbox{m/s} ~\sqrt{\frac{n}{10^{20}\mbox{m}^{-3}} } \left(\frac{10\mbox{keV}}{T}\right)^{3/2}.
\end{eqnarray}
When the ideal evolution velocity of the magnetic field lines $u_\bot$ is greater than $u_c$, which appears to be the case of primary interest in ITER, the electron skin depth is the distinguishability distance.  Since the  spatial scale of the distance between low order rational surfaces will be approximately a half meter in ITER, the ratio $\Delta_{max}/\Delta_{min}$ must be of order $10^3\approx e^7$.

The electron skin depth appears to determine $\Delta_d$ in the solar corona as well as in large tokamaks.  In the corona, the density is of order $n\sim 10^{15}~$m$^{-3}$ and the temperature is of order 100~eV, so $u_c\sim3$~m/s.   Flows of order 100~m/s tangential to the solar photosphere are observed \cite{Photospheric flows:2018}.  The ratio of the a typical distance scale of a thousand kilometers to the electron skin depth is of order $10^9\approx e^{21}$.


\section{Discussion \label{Implications}}

\subsection{New results}

Two new results are demonstrated in this paper:

(1)  Section \ref{sec:Ideal} shows that in a torus the ideal evolution of a perturbation that has an increasing magnitude leads to an ever more complicated corrugation of the magnetic surfaces.   As the amplitude of the perturbation is increased, the number of rational surfaces that must be carefully resolved increases without limit.   Except for the nested circular surfaces of cylindrical symmetry, the ratio of separations between neighboring magnetic surfaces $\Delta_{max}/\Delta_{min}$ is greater than unity, and the ratio increases the more contorted the magnetic surfaces become.  The surface contortion becomes extreme when the amplitude of the surface displacement becomes comparable to the distance between low-order rational surfaces.  An important question for any MHD code is how large can $\Delta_{max}/\Delta_{min}$ be while preserving magnetic surfaces.  For rapid magnetic surface breaking in the core of ITER, $\Delta_{max}/\Delta_{min}$ may need to be in the thousands.  $\Delta_{max}$ should be of order $a_r$, which is the reconnection scale that causes static equilibrium to be lost, while $\Delta_{min}$ should be of order $\Delta_d$, which is the spatial scale over which the distinguishability of magnetic field lines is lost.

(2) Section \ref{B-torus} shows that a single magnetic field line trajectory $\vec{x}(\varphi)$ at a given point in time determines (a) the $\Delta_{max}/\Delta_{min}$ ratio for the separation of the neighboring magnetic surfaces, (b) both the location and the localization on the magnetic surface of the place where $\Delta_{min}$ is located, and (c) whether the magnetic surface continues to exist.  When a magnetic surface exists, the curve formed by $\vec{x}(2\pi k)$ with $k$ an integer gives the cross section of the magnetic surface in the $\varphi=0$ plane.  When a magnetic surface does not exist, the points $\vec{x}(2\pi k)$ with $k$ an integer will fill an area and do not form a well defined curve.


\subsection{Importance}

The larger the tokamak, the more damage disruptions can do to the machine.  The potential for damage is severe for ITER and would be even worse in a tokamak power plant.  Empirical disruption studies become too dangerous in large tokamaks, which implies reliable numerical simulations of planned operational conditions are required.  The methods developed in this paper are important for three reasons:

(1) They provide a method for assessing the adequacy of codes used to study disruptions, not only in operating tokamaks, but also in ITER and in tokamak power plants.  The temporal and spatial resolution of existing simulations are often too low to realistically follow disruption physics in operating tokamaks.  Future tokamaks will have a larger size and generally operate at a higher electron temperature, which will make obtaining an adequate resolution even more difficult.  

(2) Much can be learned from simulations with an inadequate resolution, but a deep understanding of the phenomena being simulated is required to avoid mistakes that could endanger the machine.  

For example, the breaking of magnetic field lines and surfaces occurs at specific locations, the places where $\Delta_{min}$ is smallest.  This may have important implications for the mixing of cold impurity ions, which move extremely slowly both along and across field lines.  That is, the impurity mixing is closely related to the mixing of field line segments produced by reconnection.  $\Delta_{min}$ presumably becomes extremely localized on the surfaces as $\Delta_{max}/\Delta_{min}\rightarrow\infty$.  The extent to which existing simulation capabilities can determine this localization is unknown.

(3) Traditional theories of magnetic reconnection posit current sheets in which the current density along the magnetic field reaches a value inversely proportional to the non-ideal effects, $j_{||}\propto a_r/\Delta_d$.  The electric field parallel to the magnetic field associated with these current sheets could greatly exacerbate the already difficult problem of the plasma current being transferred from near-thermal to relativistic electrons through the runaway phenomenon.   As explained in \cite{Rec-example} and in Appendix \ref{sec:reconnection}, when the timescale to reconnection depends only logarithmically on non-ideal effects, the current density is only enhanced logarithmically, $j_{||}\propto \ln(a_r/\Delta_d)$, not inversely with the strength of those effects.  

The electron acceleration question could be subtle because particle acceleration can occur even in an ideal evolution \cite{Boozer:acc}.  Non-ideality can be represented in Faraday's law, Equation (\ref{B-ev}), by $\mathcal{E}$, Equation (\ref{EMF}). $\mathcal{E}$  is equivalent to a loop voltage, the energy gained by an electron from $\mathcal{E}$ per toroidal circuit.   The large number of toroidal circuits required in ITER to change the energy of an electron by 0.5~Mev, of order a thousand, tends to average the terms in the acceleration to zero, other than the loop voltage, when magnetic surfaces exist. 

As discussed in \cite{L/R}, a high current density will arise in ITER in a thin layer, just inside the region of intact magnetic surfaces, where an inner region of intact surfaces comes into contact with an outer region of chaotic magnetic field lines created by the breaking of magnetic surface.  The reason is that $j_{||}/B$ quickly becomes independent of position in chaotic field-line regions \cite{Boozer:j-||}.  The flat current profile in the chaotic region changes the poloidal flux between the highly conducting ITER walls and boundary between broken and intact surfaces.  However, as long as the electrical conductivity of the intact surfaces will not allow a fast change in the poloidal flux to occur; a shielding current will be driven.  A large current density means a large loop voltage which can exacerbate the runaway electron problem.

What is clear is that codes will need to accurately simulate effects even more delicate than the current spike before they can be relied upon for the safety of the ITER device or tokamak power plants.


\subsection{Additional studies}

The simplest relevant study is the time development of ideal perturbations in helical symmetry.  Only the linear limit, which fails after a sufficient number of Alfv\'en transit times, and the non-linear steady-state limit have been studied.  The the time-dependent helically symmetric calculation is not only simplified by having only two non-trivial coordinates but also because there is only one resonant rational surface, which has a known location, so the radial grid can be packed in that region.  In addition to the ideal evolution, it is also important to study of the effect of the electron skin depth, $c/\omega_{pe}$, when it is significantly larger than the radial grid and with resistivity kept at a sufficiently low level to be negligible.  

More challenging simulations are required to determine the observational effects of the breakup of magnetic surfaces.  Since magnetic field lines are defined at points in time, the trajectory of a magnetic field line that passes through a particular point in space can undergo an arbitrarily large trajectory change between time $t$ and $t+\delta t$ due to a reconnection event, even as $\delta t\rightarrow0$.  This may seem to be physical nonsense until a distinction is made between the reconnections themselves, which means topology changes, and reconnection effects, which are physical effects that result from the changes in field-line topology.  In hot, multi-kilovolt tokamak plasmas, the fastest observational effect of reconnection \cite{Boozer:j-||} is generally the spreading of energetic electrons along the reconnected magnetic field lines by collisionless streaming.  This has been observed on DIII-D to occur on a 50~$\mu$s time scale,  \cite{Paz-Soldan:2020}.  The relaxation of the parallel current density, or more precisely $j_{||}/B$, occurs \cite{Boozer:j-||} on the timescale for a shear Alfv\'en wave to propagate along the reconnected field lines.  Both empirical and theoretical evidence \cite{Boozer:j-||} imply that of order a hundred toroidal transits are required for a single magnetic field line to go from the central region to the plasma edge after a large scale breaking of magnetic surfaces has occurred in a tokamak disruption.  Effects that involve the loss of static force balance across the magnetic field lines occur on a much faster timescale, the timescale for a compressional Alfv\'en wave to propagate across the magnetic field lines.  This propagation is far faster than the most obvious causes of loss of force balance across the lines, which are adjustments in the plasma pressure and the relaxation of $j_{||}/B$ to a  constant in spatial regions covered by a single field line.  Consequently, the plasma generally remains in force balance across the magnetic field as the electron energy and $j_{||}/B$ relax along the magnetic field.

\subsection{Relation to other reconnection theories}

\subsubsection{Traditional model of magnetic reconnection}
 
 The traditional model of magnetic reconnection  \cite{Schindler:1988} is steady state \cite{Amitava:2004}  with a sharp boundary between two regions in which a component of the magnetic field has the same magnitude but opposite signs.  Figure \ref{fig:plasmoid}  illustrates the traditional model in its modern plasmoid form \cite{plasmoid}.  The entire reconnected flux is processed through the thin boundary layer between the two flux regions in which the current density becomes singular as the distinguishability distance goes to zero.  
 
 The traditional model of reconnection presents a distinct picture from that given by the mathematical form of a generic ideal evolution \cite{Boozer:rec-phys}, Appendix \ref{sec:reconnection}.  In a generic ideal evolution, magnetic flux tubes becomes exponentially distorted, which means the ratio of separation between the closest and furtherest separation between pairs of field lines $\Delta_{max}/\Delta_{min}$ becomes exponentially large.  Rapid reconnection arises when that ratio becomes comparable to $a_r/\Delta_d$, the spatial scale $a_r$ over which reconnection can cause a loss of static equilibrium divided by the distinguishability distance.  Although $\Delta_{max}/\Delta_{min}$ becomes extremely large, the current density does not, \cite{Boozer:separation} and Appendix \ref{sec:reconnection}, even though it flows in narrow sheets.  A simple corona-like model of an evolving magnetic field \cite{Rec-example} illustrates these generic features of an ideal evolution. 
 
 As discussed in Appendix \ref{boundary-conditions}, boundary conditions in plasmoid theory have problematic features.
 
 \subsubsection{Turbulence models}
 
 Turbulence can enhance the rate of reconnection.  As discussed in Section VIII of \cite{Rec-example}, turbulence is of two types, (1) an enhanced effective resistivity for $j_{||}$, which would make $\Delta_d$ larger and  (2) a turbulent velocity of magnetic field lines.  
 
 The $j_{||}$ current densities that are expected, \cite{Boozer:separation} and Appendix \ref{sec:reconnection}, even when the magnetic surfaces are highly contorted, are not extremely high compared to those that commonly arise in tokamaks, and an enhanced effective resistivity for $j_{||}$ is not noted in tokamak experiments.  
 
 There is an extensive literature on the enhancement of reconnection by turbulence in the magnetic field line velocity \cite{Lazarian:1999,Eyink:2011,Eyink:2015,Matthaeus:2015,Matthaeus:2020}, which is reviewed in \cite{Lazarian:2020rev}.  Turbulence in the ideal field line velocity would make the surfaces even more contorted, but could not directly break the surfaces.  If MHD codes can accurately follow the evolution of an ideal perturbation, they should show the enhanced surface distortion produced by ideal instabilities.  As long as surfaces exist, the methods of analysis developed in this paper remain valid.


\subsubsection{Models by Eric Priest}

Eric Priest has inspired a large body of work on three-dimensional structures that tend to concentrate currents and thereby lead to enhanced reconnection \cite{Priest:2016}.  These structures are null points, where the magnetic field vanishes, separators, which are magnetic field lines joining null points, and quasi-separatrix layers, which are regions in a magnetic field where the gradient of a footpoint mapping is large.

The relation of these topics to the rapid breakup of tokamak magnetic surfaces is obscure.  Quasi-separatrix layers are related to papers by Boozer and Elder \cite{Rec-example} and by Reid, Parnell, Hood, and Browning \cite{Reid:2020}, which discuss two closely related models of reconnection in the solar corona.

\section*{Acknowledgements}
This work was supported by the U.S. Department of Energy, Office of Science, Office of Fusion Energy Sciences under Award Numbers DE-FG02-95ER54333, DE-FG02-03ER54696, DE-SC0018424, and DE-SC0019479.

\section*{Author Declarations}

The author has no conflicts to disclose.


\section*{Data availability statement}

Data sharing is not applicable to this article as no new data were created or analyzed in this study.


\appendix


 
 \section{Mathematical properties of Faraday's Law \label{sec:reconnection} }
 
 Any magnetic evolution is described by Farday's Law, $\partial \vec{B}/\partial t=-\vec{\nabla}\times\vec{E}$.  This equation and $\vec{\nabla}\cdot\vec{B}=0$ have subtle mathematical properties that clarify and constrain the phenomenon of magnetic reconnection.  A number of these properties was demonstrated decades ago in highly-cited papers.  Nevertheless, they and their importance are not widely appreciated in the plasma community.  Their appreciation is needed to understand this paper and to place it in context, so this appendix will review and sketch proofs of these properties.
 
 
 \subsection{Representation of the electric field}
 
 Mathematics implies an arbitrary vector $\vec{E}(\vec{x})$ in three-space can be represented in terms of another vector $\vec{B}(\vec{x})$ that has no zeros in the region of interest,
\begin{equation}
\vec{E} = -\vec{u}\times\vec{B} -\vec{\nabla}\Phi +\mathcal{E}\vec{\nabla}\ell.  \label{E rep}
\end{equation}
$\Phi$ is a single valued potential, and $\ell$ is the distance along the vector $\vec{B}$.  This means the field lines of $\vec{B}$ are given by $d\vec{x}/d\ell = \vec{B}/B\equiv\hat{b}$ at a given point in time; the vector $d\vec{\ell} \equiv\hat{b} d\ell$.  The proof of Equation (\ref{E rep}) is simple.  The component of $\vec{E}$ along $\vec{B}$ gives $\hat{b}\cdot\vec{E}= -\partial{\Phi}/\partial \ell +\mathcal{E}$, where $\mathcal{E}$ is a constant along $\vec{B}$, which must be chosen to make $\Phi$ single-valued.  It is essentially the concept of an electromotive force in electrodynamics,  
\begin{equation}
 \mathcal{E}=\frac{\int_{-L/2}^{L/2} \vec{E}\cdot d\vec{\ell}}{L}.  \label{EMF}
 \end{equation}
 When $\vec{B}$ lies on toroidal surfaces, the integration distance $L$ goes to infinity and $2\pi R_0\mathcal{E}$ is the loop voltage with $R_0$ a suitably averaged major radius of the surface.   The components of $\vec{E}$ perpendicular to $\vec{B}$ determine $\vec{u}_\bot$, which are the two components of $\vec{u}$ that are perpendicular to $\vec{B}$.
 
As discussed in Section IV.C of \cite{Boozer:null-X}, the integral of Equation (\ref{EMF}) is bounded by an infinitesimal sphere around a field null, where $\vec{B}=0$, if one exists along field line, and by specified boundaries, such as a perfectly conducting wall.

 This form for a general electric field and its derivation date back forty years to Section IV of a paper with 302 citations \cite{Boozer:coordinates}.  Although this paper is highly cited---it showed the existence of what are now called Boozer coordinates---it is apparently little read since not only is its form for $\vec{E}$ ignored but also the stated purpose.  That purpose was to show that asymmetric scalar-pressure equilibria can have finite transport despite Grad's conjecture \cite{Grad:1967}.
 
 The use of Equation (\ref{E rep}) for the electric field is sometimes questioned.  How is it possible for a simple calculation to remove explicit consideration of the Hall and the viscosity terms of Ohm's law, which are important in the reconnection literature?  The answer is that these terms primarily determine the difference between the magnetic field line velocity $\vec{u}_\bot$ and the plasma velocity $\vec{v}$.  Equation (\ref{E rep}) shows the velocity $\vec{v}$ affects magnetic reconnection only indirectly.  The non-relativistic Lorentz transformation says that $\vec{E}+\vec{u}\times\vec{B}$ is the electric field in the frame moving with the velocity $\vec{u}$;  Equation (\ref{E rep}) gives the form of the electric field in that frame.  The traditional Ohm's law gives the electric field in the plasma frame.  A change in reference frame can demonstrate that a problem is much simpler than one might think.  A familiar example is the center-of-mass reference frame in classical mechanics, which reduces the two-body problem to a one-body problem.
 
 Using Equation (\ref{E rep}) to represent the electric field, Faraday's law becomes
 \begin{equation} 
 \frac{\partial\vec{B}}{\partial t} = \vec{\nabla}\times\left( \vec{u}_\bot\times\vec{B}-\mathcal{E}\vec{\nabla}\ell\right) \label{B-ev} 
 \end{equation}
 since $\vec{\nabla}\times(\vec{\nabla}\Phi)=0$. 
 
\subsection{Ideal magnetic evolution and the field line velocity}

More than sixty years ago, in 1958, Newcomb proved \cite{Newcomb} that when $\mathcal{E}=0$, magnetic field lines move with the velocity $\vec{u}_\bot$.   The proof is simple.  Write the magnetic field in the well known Clebsch form $2\pi\vec{B}=\vec{\nabla}\psi\times\vec{\nabla}\Theta$ and show the evolution of field-line labels $\psi$ and $\Theta$ can be written as convective derivatives,  $\partial\psi/\partial t+\vec{u}_\bot\cdot\vec{\nabla}\psi=0$ and $\partial\Theta/\partial t+\vec{u}_\bot\cdot\vec{\nabla}\Theta=0$.  When $\mathcal{E}\neq0$, one can show $\Delta_d$ is non-zero and reconnection occurs  \cite{Boozer:rec-phys}.

Although Newcomb's paper has had 201 citations, important points that he made in the abstract are not generally recognized within the plasma physics community.  (1) The flow speed of the magnetic field lines is distinct from that of the plasma.  (2) The magnetic flux enclosed by a curve that moves with field line velocity in an ideal evolution, $\mathcal{E}=0$, is conserved, but the flux may fail to be conserved when the motion is defined by the plasma velocity.  (3)  It is impossible to have the flux conserved when the curve moves with the plasma velocity when it is not conserved with the field line velocity.

When applied to toroidal plasmas, Newcomb's ideal-evolution proof for $\mathcal{E}=0$  was always understood to imply a perfect preservation of  magnetic surfaces including the toroidal magnetic flux enclosed by each surface and the twist of the field lines in each surface, the rotational transform $\iota$.  An explicit proof for toroidal surfaces is given in \cite{Boozer:RMP}. 
  

\subsection{Chaotic flows}

The concept of a chaotic flow, $\vec{u}(\vec{x},t)$,  comes from a branch of mathematics, chaos theory.  In a chaotic flow, pairs of neighboring streamlines $\vec{x}(t)$ of a flow,  $d\vec{x}/dt=\vec{u}$, separate exponentially over time with such pairs existing through a non-zero volume.  Neighboring means having an infinitesimal spatial separation.

Although chaotic flows are defined by an exponentially increasing separation of infinitesimally separated streamlines, the exponentiation characterizes the behavior of pairs of streamlines as long as their separation is small compared to the spatial scale of the flow $\vec{u}(\vec{x},t)$.  In the limit of larger separations, the separation tends to  increase  only diffusively; in simple cases this means as $\sqrt{t}$. 
  
The remarkable fact is that chaotic flows are essentially universal even for divergence-free flows in two spatial dimensions, $dx/dt=-\partial h/\partial y$ and $dy/dt=\partial h/\partial x$ when the stream function, or Hamiltonian, $h(x,y,t)$ depends on time $t$ as well as $x$ and $y$.  Chaos in Hamiltonian systems was considered sufficiently remarkable in 1986 to justify a review \cite{Lighthill:1986}, but chaotic flows are universally recognized to be the cause of enhanced mixing in near-ideal fluids after the 1984 paper of Aref \cite{Aref:1984,Aref:2017}.  For example, the equilibration of temperature in a room requires of order ten minutes instead of the two weeks expected from diffusion alone.   Thermal equilibration and magnetic reconnection are related examples of chaotic flows causing non-ideal relaxations that depend only logarithmically on non-ideal effects \cite{Boozer:rec-phys}.

\subsection{Exponential separation of magnetic field lines}

 At a particular point in time,  two neighboring magnetic field lines have a minimum $\Delta_{min}$ and a maximum $\Delta_{max}$ separation, where neighboring implies the limit $\Delta_{max}\rightarrow0$.   As an ideal evolution proceeds, the ratio $\Delta_{max}/\Delta_{min}$ for a particular pair of lines can increase.  Magnetic reconnection becomes important when $\Delta_{max}/\Delta_{min}\gtrsim a_r/\Delta_d$ where $a_r$ is the spatial scale at which reconnection causes a lost of static equilibrium and $\Delta_d$ is the distinguishability distance.
 
 Magnetic field line chaos is usually defined by pairs of field lines having a non-zero Lyapunov exponent, $\gamma_\infty$, throughout a finite volume.  The field line exponentiation of a pair of neighboring pair of lines over a field line integration of length $L$ is defined by
 \begin{eqnarray}
e^{\sigma(L)}&\equiv& \frac{\Delta_{max}}{\Delta_{min}}.
 \end{eqnarray}
 The Lyapunov exponent is defined as
  \begin{eqnarray}
 \gamma_\infty &\equiv& \lim_{L\rightarrow\infty}\frac{\sigma(L)}{L}.
 \end{eqnarray} 
 
 Since magnetic reconnection occurs when $\Delta_{max}/\Delta_{min}\approx a_r/\Delta_d$, the field lines need not be chaotic for an increasing exponentiation to cause reconnection to occur on a timescale determined by the ideal evolution.  Only $\sigma(L)\approx\ln(a_r/\Delta_d)$ is required.  Reconnection occurs in a solar-corona model \cite{Rec-example} on a time scale set by the ideal evolution with the length $L$ finite.  Consequently, the field-line Lyapunov exponent $\gamma_\infty$ is not properly defined.  When the magnetic field lines lie on magnetic surfaces in a torus, the field-line Lyapunov exponent is well defined but zero.  Nevertheless, an increasing $\Delta_{max}/\Delta_{min}$ can produce reconnection on a timescale determined by the ideal evolution.

 Although $\sigma(L)$ is defined for particular pairs of neighboring field lines, it represents the distance field lines can exponentially separate when $\Delta_{max}\lesssim a_u$, where $a_u$ is the characteristic spatial scale across $\vec{B}$ on which the flow $\vec{u}_\bot$ varies. This is illustrated in Figure 3c of \cite{Rec-example}. At any point in time, $\sigma(L)$ is an extremely complicated function of which pair of field lines is chosen; the spatial gradient of $\sigma(L)$ increases exponentially.
 
 
 \subsection{Current density when $B$-lines exponentially separate}
 
 A number general properties of magnetic field lines can be obtained from the equations for magnetic field lines infinitesimally separated from an arbitrarily chosen line \cite{Boozer:separation}, which is called the central line.  The analysis is made using the Courant and Snyder \cite{Courant-Snyder} intrinsic coordinates $(\rho,\alpha,\ell)$ about the space curve defined by the arbitrarily chosen central line.  The intrinsic coordinates locally resemble cylindrical coordinates with $\rho$ the radius, $\alpha$ an angle relative to the curvature of the space curve, and $\ell$ the distance along the curve.  The neighboring field lines are given by a Hamiltonian \cite{Boozer:separation}
 \begin{eqnarray}
 H(\tilde{\psi},\alpha,\ell)&=&\tilde{\psi} h(\alpha,\ell),  \mbox{   where  } \\
h &=& k_\omega(\ell) + k_q(\ell) \cos\big(2\alpha -\varphi_q(\ell)\big), \hspace{0.2in}\\
k_\omega &=& \tau(\ell) +\frac{\mu_0 j_{||}(\ell)}{2B_0(\ell)},  \mbox{   and   }\\
\tilde{\psi} &\equiv& \frac{B_0(\ell)}{2} \rho^2,  \mbox{   so $2\pi\tilde{\psi}$ is a flux;   }
 \end{eqnarray}
$k_q$ and $\varphi_q$ are the amplitude and phase of the quadrupole moment of the magnetic field expanded around the central line.  Hamilton's equations are $d\alpha/d\ell = \partial H/\partial\tilde{\psi}$ and $d\tilde{\psi}/d\ell=-\partial H/\partial\alpha$.  The expression for $k_{\omega}$ uses Ampere's Law.

The field lines given by this Hamiltonian have fundamentally different properties depending on whether it  has a dependence on $\ell$.  When $H$ depends on $\ell$, a circular flux tube at $\ell=0$ can become elliptical with a ratio of semimajor axis to semiminor axis equal to $e^{2\sigma(\ell)}$; the exponentiation, $\sigma(\ell)$, can become arbitrarily large even when the Lyapunov exponent, $\lim_{\ell\rightarrow\infty} \sigma(\ell)/\ell=0$.

The parallel current density $j_{||}(\ell)$ appears directly in the Hamiltonian, and the parallel current density required to produce the quadrupole field can be estimated.  The answer is that $j_{||}/B$ scales linearly with the exponentiation $\sigma$.  As has been seen, magnetic reconnection becomes important when $\sigma \gtrsim \ln(a_r/\Delta_d)$, so the result that the current density required to obtain reconnection scales as $\ln(a_r/\Delta_d)$ as $a_r/\Delta_d\rightarrow0$ and not as $a/\Delta_d$ as it does in two dimensional reconnection theory \cite{Schindler:1988}, which is equivalent to $H$ being independent of $\ell$.



\subsection{Boundary conditions \label{boundary-conditions}}

 The mathematics of partial differential equations shows that appropriate boundary conditions are essential for the solution of Equation (\ref{B-ev}) for a magnetic evolution.  The simplest well-posed boundary condition is to surround the entire reconnecting region by a perfect conductor, which can be assumed to move or deform in time in order to drive the evolution.  In principle, one can have outgoing boundary conditions on an enclosing surface, but such boundary conditions are subtle.  What differences are produced by field lines that preserve their connections even after they have crossed the enclosing surface versus those that do not?
 
 The effects of boundary conditions are both subtle \cite{Boozer:space} and rarely emphasized in traditional reconnection models, Figure \ref{fig:plasmoid}. If the $x$ and $y$ directions of this figure are taken to be periodic, then the system is a mathematically toroidal; the discussion of Section \ref{sec:B-torus} applies directly.  Tearing requires the magnetic field lines that pass through the $z=0$ surface close on themselves, the equivalent of a rational surface.  When the $y$ direction, which is perpendicular to the page in Figure \ref{fig:plasmoid}, is unbounded, the interpretation of figure and of the concept of tearing is obscure. Outgoing boundary conditions are required unless the central line is a $\vec{B}=0$ line, which is mathematically unstable.  An infinitesimal perturbation can transform a line null into well separated point nulls.

 
  \subsection{Energy exchange}
 
 Using Poynting's theorem and Ampere's law, one can derive the ideal evolution equation for the magnetic energy density  \cite{Boozer:rec-phys}:
  \begin{eqnarray}
   &&\frac{\partial}{\partial t}\left(\frac{B^2}{2\mu_0}\right) +\vec{\nabla}\cdot\left(\frac{B^2}{2\mu_0} \vec{u}_\bot\right)  \nonumber\\
   && \hspace{0.5in} =-\left(\frac{B^2}{2\mu_0}\right)\big(\vec{\nabla}\cdot\vec{u}_\bot+2\vec{u}_\bot\cdot\vec{\kappa}\big), \label{B-energy} \hspace{0.2in}
  \end{eqnarray}
where $\vec{\kappa}\equiv\hat{b}\cdot\vec{\nabla}\hat{b}$ is the curvature of the magnetic field lines.  When Equation (\ref{B-energy}) is integrated over the volume enclosed by a moving perfect conductor, the left-hand side gives the energy input due to the motion of that perfect conductor.  The right-hand side implies that only a field-line velocity that satisfies $\vec{\nabla}\cdot\vec{u}_\bot+2\vec{u}_\bot\cdot\vec{\kappa}=0$ can flow without large energy exchanges with the plasma in which the field is embedded.  Three spatial coordinates are required to satisfy this condition during the evolution toward a rapid reconnection.  Two coordinates perpendicular to $\vec{B}$ are required to satisfy $\vec{\nabla}\cdot\vec{u}_\bot+2\vec{u}_\bot\cdot\vec{\kappa}=0$ and a third coordinate, which gives the variation along $\vec{B}$, is required for a non-trivial magnetic reconnection problem.
  
When the magnetic evolution is in a two-dimensional space, an exponential increase in the separation of magnetic field lines cannot be responsible of fast magnetic reconnection in traditional problems, for that would require an exponential increase in the magnetic energy. Schindler, Hesse, and Birn \cite{Schindler:1988} proved that  in two-dimensional space reconnection that competes with evolution requires a thin reconnection layer in which the current density reaches $j\approx B_{rec}/\mu_0\Delta_d$, where $B_{rec}$ is the reconnecting magnetic field and $\Delta_d=\eta/\mu_0u_\bot$ is the resistive distinguishability distance.   The ratio of the cross magnetic-field scale $a_B$ to the resistive distinguishability scale is the magnetic Reynolds number $R_m =\mu_0 u_\bot a_B/\eta$. 

 
 \subsection{General ideal evolution of $\vec{B}$ \label{sec:gen-ev} }
 
The magnetic field $\vec{B}(\vec{x},t)$ is completely defined in an ideal evolution by the magnetic field line flow $\vec{u}_\bot$ and the initial magnetic field, $\vec{B}_0(\vec{x},t=0)$.  The relation has a long history, which was reviewed by Stern  in a paper with ninety citations \cite{Stern:1966}.  The relation uses the Jacobian matrix $\tensor{\mathcal{J}}$ of Lagrangian coordinates $\vec{x}(\vec{x}_0,t)$.

This relation between $\vec{B}(\vec{x},t)$ and $\vec{B}_0(\vec{x},t=0)$ is important because simple properties of an arbitrary ideal evolution determine the evolution of the magnetic field energy density, $B^2/2\mu_0$, the separation between neighboring magnetic field lines, and the current density throughout space.  Again one finds \cite{Boozer:rec-phys} that a large increase in the separation between magnetic field lines $e^{\sigma(\ell,t)}$ with little change in the magnetic energy density gives a current density that scales a $\sigma$.

The definition of Langrangian coordinates and their Jacobian matrix starts with the position vector  at $t=0$ of Cartesian coordinates, $\vec{x}_0=x_0\hat{x} +y_0\hat{y} + z_0\hat{z}$.  The coordinates defined by $\vec{x}_0$ become Lagrangian coordinates if the position vector at time $t$ is $\vec{x}(\vec{x}_0,t)=x(x_0,y_0,z_0,t)\hat{x} +y(x_0,y_0,z_0,t)\hat{y} + z(x_0,y_0,z_0,t)\hat{z}$, where
\begin{equation}
\frac{\partial\vec{x}(\vec{x}_0,t)}{\partial t} \equiv  \vec{u}_\bot(\vec{x},t). 
\end{equation} 
Jacobian matrix of Langrangian coordinates is 
\begin{equation}
\tensor{\mathcal{J}}(\vec{x}_0,t) \equiv \frac{  \partial \vec{x}(\vec{x}_0,t)  }{ \partial \vec{x}_0}.
\end{equation}
The determinant of $\tensor{\mathcal{J}}$ is the Jacobian $\mathcal{J}$ of Lagrangian coordinates.

The magnetic field at time $t$ is then
\begin{equation}
\vec{B}(\vec{x},t) = \frac{ \tensor{\mathcal{J}} }{{\mathcal{J}}}\cdot\vec{B}_0(\vec{x}_0). \label{gen-ev}
\end{equation}

Mathematical implications of this equation, and hence any ideal evolution, can be obtained using a Singular Value Decomposition (SVD) of $\tensor{\mathcal{J}}$, which yields three eigendirections and three Singular Values at each point in space.  Singular Values are positive, real numbers, so they can always be ordered by their magnitude---large, middle, and small---although all equal unity at $t=0$.

The solution to Equation (\ref{gen-ev}) becomes highly constrained \cite{Boozer:rec-phys} in regions in which the largest Singular Value becomes large, $e^{\sigma(\vec{x},t)}$ with $\sigma >>1$.  Assuming $B^2$ does not exponentially increase, the magnetic field must lie in the eigendirection of the middle singular value and magnetic flux tubes become exponentially distorted.  The current density lies in ribbons along the magnetic field.  These ribbons become exponentially, $e^\sigma$, broad in the eigendirection of the large singular value and exponentially thin, $e^{-\sigma}$, in the eigendirection of the small singular value.  Since these features are universal, they can be illustrated by the use of even a simple model \cite{Rec-example}.


\section{Effect of two-dimensional ideal magnetic perturbations \label{2D pert}}

The most important perturbations to nested toroidal magnetic surfaces are resonant.  A resonant perturbation has a normal magnetic field to the unperturbed surfaces which has a least one Fourier coefficient with a poloidal mode number $M$ and toroidal mode number $N$, that is resonant  with the rotational transform $\iota(\psi_{r})=N/M$ on a magnetic surface $\psi=\psi_r$ within the region of nested toroidal surfaces.  

When a toroidal plasma is perturbed---even a plasma that was axisymmetric---there are no symmetry directions.  An implication is that there is no single-term $(M,N)$ Fourier decomposition of the perturbing normal field within the plasma volume, and multiple rational surfaces are generally affected.  The interactions associated with the various Fourier terms is required to obtain an extremely large ratio, $\Delta_{max}/\Delta_{min}$, in the separation of neighboring magnetic field lines throughout a volume and not just on a surface.  It is the volumetric effect that leads to large scale magnetic reconnection.

Unfortunately, detailed studies of ideal perturbations have only been carried out with the assumption that the system has perfect helical symmetry in a limit in which the curvature of the magnetic surfaces can be ignored so that Cartesian coordinates can be employed.  It is also assumed that the plasma pressure is zero.  Steady-state studies have been carried out in \cite{Boozer-Pomphrey,Zhou:2016,Zhou:2019,Huang:2021} and are based on a method developed by Rosenbluth, Dagazian, and Rutherford \cite{RDR-kink} to study the $M=1$ $N=1$ kink in tokamaks.

\subsection{Cartesian coordinate approximation}

In helical symmetry, any perturbation is a periodic function of $M\theta - N\varphi$ in $(r,\theta,z)$ cylindrical coordinates, where $\varphi\equiv z/R_0$ with $2\pi R_0$ the wavelength of the perturbation in the $z$ direction.  The helical angle $\theta_h$ and the helical symmetry vector $\vec{h}$ are
\begin{eqnarray}
&& \theta_h \equiv M\theta - N\varphi \mbox{   and   } \\
&& \vec{h} \equiv \frac{M^2\hat{z}+\frac{r}{R_0} MN\hat{\theta}}{M^2 + N^2\frac{r^2}{R_0^2}},  \mbox{   where    } \\
&& \vec{\nabla}\cdot\vec{h}=0;\\
&& \vec{h}\cdot\vec{\nabla}\theta_h=0\\
&& \vec{\nabla}\times \vec{h} = \frac{2}{R_0}\frac{MN}{M^2 + N^2\frac{r^2}{R_0^2}}\vec{h}.
\end{eqnarray}

The magnetic field has helical symmetry when
\begin{eqnarray}
&& \vec{B} = B_0\vec{h}+ \vec{\nabla}\times(A(r,\theta_h) \vec{h}); \label{helical B}\\
&& \vec{B}\cdot\vec{\nabla}A = 0
\end{eqnarray}
The magnetic field lines lie in surfaces of constant $A$, where $A\vec{h}$ is the vector potential in the direction of helical symmetry.

To simplify the problem further, it is assumed that $R_0\rightarrow\infty$.  In this limit, $\vec{h}=\hat{z}$.  The current density is $\mu_0 \vec{j}=\vec{\nabla}\times(\vec{\nabla}\times A\vec{h})$ with $\vec{h}\rightarrow \hat{z}$, so 
\begin{equation}
\mu_0 j_z = - \nabla^2 A. \label{j_z}
\end{equation}

The operator $\vec{B}\cdot\vec{\nabla}f$ is of particular importance.  When the function $f(\vec{x})$ is helically symmetric, $\vec{h}\cdot\vec{\nabla}f=0$, and
\begin{eqnarray} 
\vec{B}\cdot\vec{\nabla}f &=&- \vec{h}\cdot (\vec{\nabla}A\times\vec{\nabla}f);\\
&=&\vec{B}\cdot\vec{\nabla}\theta_h \frac{\partial f(A,\theta_h)}{\partial\theta_h}.
\end{eqnarray}
In the unperturbed state, $\theta_h = M\theta - N\varphi$, so $\vec{B}\cdot\vec{\nabla}\theta_h=(\iota(A) M-N) \vec{B}\cdot\vec{\nabla}\varphi$, where $\vec{B}\cdot\vec{\nabla}\varphi=B_0/R_0$, which is a constant as $R_0\rightarrow\infty$.  Consequently, the operator
\begin{eqnarray}
\frac{\vec{B}\cdot\vec{\nabla}f}{B_0} &=&\frac{\iota(A) M-N}{R_0} \frac{\partial f(A,\theta_h)}{\partial\theta_h} \label{k_||}
\end{eqnarray}
in the limit as the perturbation goes to zero.

The unperturbed vector potential near the rational surface $\iota(r_s)=N/M$ is
\begin{eqnarray}
A_0 &=& - \frac{\mu_0 \bar{J}_0}{2} r_s^2\left( \ln(r/r_s) - \frac{r-r_s}{r_s}\right) \label{A_0 term}\\
&\approx& \frac{\mu_0 \bar{J}_0}{4}x_0^2 +\cdots, \mbox{   where }  \label{A_0 form}\\
x_0 &\equiv& r-r_s
\end{eqnarray}
and $\bar{J}_0$ is the average current density in the region enclosed by the rational surface.  For both simplicity and to be definite, the current density in the unperturbed plasma is assumed to be zero at the rational surface. The term involving $(r-r_s)/r_s$ in Equation (\ref{A_0 term}) comes from the unperturbed field on the rational surface being equal to $\vec{B}(r_s) = B_0\vec{h} \approx B_0(\hat{z} + (Nr_s/MR_0) \hat{\theta})$.  The radial derivative of the rotational transform at the rational surface is
\begin{eqnarray}
\iota' \equiv -\frac{\mu_0\bar{J}_0R_0}{2B_0r_s}
\end{eqnarray}
The rotational transform is order unity, so the ratio $B_0/R_0$ must remain constant as the limit $R_0\rightarrow\infty$ is taken.

An ideal perturbation can move the magnetic surfaces, which remain surfaces of constant-$A$.  The implication is that if a magnetic surface is perturbed from being independent to being dependent on $\theta_h$, then its position is
\begin{eqnarray}
x&=&x_0+\xi(x_0,\theta_h) \mbox{   where    } \\
A_0(x_0) &=& \frac{\mu_0 \bar{J}_0}{4}x_0^2 \label{A0}
\end{eqnarray}
and $\xi(x_0,\theta_h)$ is the displacement of the surface.   That is, the vector potential in the presence of the perturbation is
\begin{eqnarray}
A&=& A_0(x-\xi) \\
&=& \frac{\mu_0 \bar{J}_0}{4} (x^2-2x\xi +\xi^2).  \label{A-xi}
\end{eqnarray}
As $x\rightarrow\infty$, the vector potential in the presence of the perturbation goes to
\begin{eqnarray}
A&\rightarrow& \frac{\mu_0 \bar{J}_0 x^2}{4}\left(1 - 2\frac{\xi}{x}\right).
\end{eqnarray}
A cosinusoidal perturbation is assumed to be driven at a large distance $x$ from the resonant surface  in the sense $\Big|x/\xi \Big|>>1$, but close to the rational surface, which has a radius $r_s$, in the sense $\Big|x/r_s \Big|<<1$.

The approximations that have been made reduce the problem to $(x,y,z)$ Cartesian coordinates, where 
 \begin{eqnarray}
 x(x_0,\theta_h) &=& x_0 + \xi(x_0,\theta_h), \\
 y(\theta_h) &=& r_s \theta_h, 
 \end{eqnarray}
 and $z$ is the direction of symmetry.  Using Equation (\ref{A0}), the unperturbed coordinate $x_0$ also specifies a particular value of the vector potential, which is unchanged by the perturbation.  The radius $r_s$ is the radius of the resonant magnetic surface. 
 
 
 \subsection{Force-free solution \label{sec:force-free} }
 
 When the plasma pressure is zero, the only longterm solution for force balance is for the current to be force free, which means $\vec{j}=(j_{||}/B)\vec{B}$, and divergence free, which implies $\vec{B}\cdot\vec{\nabla}(j_{||}/B)=0$.  Using Equation (\ref{j_z}), the infinite-time solution for the magnetic field subjected to perturbation that is time-independent is a solution of the equation 
 \begin{eqnarray}
\nabla^2 A&=& -  \mu_0 j_z(A),
\end{eqnarray}
with the displacement of $\xi$ satisfying $A(x,y)=A_0(x-\xi)$.  The boundary condition is on a surface $r=b$.  On the $r=b$ surface, the displacement is $\xi_b = - \xi_d \cos\theta_h$, or $\xi_b = - \xi_d \cos(y/r_s)$.  The solution gives not only the displacement $\xi(x,y)$ but also the self-consistent force-free current $j_z(A)$.  For simplicity, it is usually assumed the perturbation is applied on both sides of the rational surface in a balanced way, so the rational surface itself remains at its unperturbed position.

The most important results \cite{Zhou:2021} are that a magnetic surface, which was separated from the rational surface by a distance $x_0$, has a closest approach $x \propto x_0^2$ and a furthrest separation $x\propto x_0^{2/3}$ in the region in which $x_0$ is small compared to the imposed displacement.  The width of the region of furthrest separation scales as $x_0^{1/3}$, so the cross-sectional area of this region scales as $x_0$.  The magnetic field $B_y$, which is zero at the rational surface becomes a non-zero constant as the rational surface is approached.   Figure 10 in \cite{Huang:2021} demonstrates these scalings except that of the closest approach.  As discussed in Appendix \ref{sec:time}, the time that it would take $j_z$ to become a function of $A$ alone on a surface is the time it would take a shear Alfv\'en wave to cover the surface.  This is determined by how long the Alfv\'en wave takes to propagate through the region of furtherest separation, which is proportional to $1/x_0$ just as in the unperturbed system.

 
 \subsection{Time development \label{sec:time}}
 
 \subsubsection{Evolution equations in helical symmetry}
 
 The infinite-time solution has a singular current density at the rational surface, but this current grows only algebraically in time. 
 
 Two additional equations are required to describe the evolution of a magnetic field embedded in an ideal but pressureless plasma.  
 
  One additional equation describes the evolution of the vector potential, $\partial\vec{A}/\partial t=-\vec{E}$.  The ideal electric field has the form $\vec{E}=\vec{v}\times\vec{B}-\vec{\nabla}\Phi$, so the evolution of the component of the vector potential parallel to the magnetic field is
   \begin{eqnarray} 
 \frac{\partial A}{\partial t} &=& \frac{\vec{B}\cdot\vec{\nabla}\Phi}{B_0} \label{A-ev}
 \end{eqnarray}   
 The time independence of the vector potential associated with the magnetic field $B_0\hat{z}$ implies $\vec{B}\times\vec{E} =0$, which requires the velocity 
 \begin{eqnarray}
 \vec{v}&=&-\hat{z}\times\frac{\vec{\nabla}\Phi}{B_0},  \mbox{    so   } \\
 \vec{\nabla}\times\vec{v} &=& -\hat{z} \frac{\nabla^2\Phi}{B_0}.
 \end{eqnarray}

 The other additional equation is for the evolution of the velocity, which is the force-balance equation, $\rho_0\partial \vec{v}/\partial t=\vec{j}\times\vec{B}$.  The curl of this equation implies
 \begin{eqnarray}
 \rho_0\frac{\partial \vec{\nabla}\times\vec{v}}{\partial t}&=&\vec{B}\cdot\vec{\nabla}j_z, \mbox{   so   }\\
 \frac{\partial \nabla^2\Phi}{\partial t} &=&V_A^2\frac{\vec{B}\cdot\vec{\nabla}}{B_0}\nabla^2 A, \mbox{    where  }
  \label{Phi-ev}\\
  V_A^2 &\equiv& \frac{B_0^2}{\mu_0\rho},
 \end{eqnarray}
 is the Alfv\'en speed.  Equations (\ref{A-ev}) and (\ref{Phi-ev}) together with $\nabla^2A=-\mu_0j_z$ imply the current relaxes to being force-free, $j_z(A)$, on the time scale of shear Alfv\'en wave propagation.  The operator that gives parallel wavenumber of the Alfv\'en waves, $k_{||} = (\vec{B}/B_0)\cdot\vec{\nabla}$, is inversely proportional to the distance from the rational surface.
 

\subsubsection{Hahm and Kulsrud linear solution}

In 1985 Hahm and Kulsrud \cite{Hahm-Kulsrud} solved  Equations (\ref{A-ev}) and (\ref{Phi-ev}) in the limit in which the quadratic $\xi$ term can be ignored in $A= (\mu_0 \bar{J}_0/4) (x^2-2x\xi +\xi^2)$, Equation (\ref{A-xi}), and elsewhere in the analysis.  In particular, the small perturbation limit  of the operator $(\vec{B}/B_0)\cdot\vec{\nabla}$, Equation (\ref{k_||}), was used.  Although it was assumed that the distance $x$ from the rational surface was larger than $\xi$, the distance $x$ is assumed to be much smaller than the radius $r_s$ of the rational surface, so $\nabla^2 \rightarrow \partial^2/\partial x^2$.  This assumption makes the exponential increase, $\sim \exp(Mx/r_s)$ that occurs as the source of the perturbation is approached in a curl-free magnetic field unimportant.   

Equations (\ref{A-ev}) and (\ref{Phi-ev}) become
\begin{eqnarray}
\frac{\partial A}{\partial t} &=& \frac{M\iota' x}{R_0} \frac{\partial\Phi}{\partial\theta_h};\\
\frac{\partial }{\partial t}\frac{\partial^2\Phi }{\partial x^2} &=&V_A^2 \frac{M\iota' x}{R_0} \frac{\partial }{\partial \theta_h}\frac{\partial^2 A}{\partial x^2}, \mbox{ so   } \hspace{0.2in}\\
\frac{\partial^2 }{\partial t^2} \frac{\partial^2 }{\partial x^2}\frac{A}{x}&=&\left(V_A \frac{M\iota'}{R_0}\right)^2 x \frac{\partial^2 }{\partial x^2} \frac{\partial^2 }{\partial \theta_h^2}A. \label{HK-A}
\end{eqnarray}
The part of $A$ that depends on time and $\theta_h$ is proportional to $x\xi$, so Equation (\ref{HK-A}) can be written
\begin{eqnarray}
\frac{\partial^2 }{\partial t^2} \frac{\partial^2 \xi}{\partial x^2}&=&\left(V_A \frac{M\iota'}{R_0}\right)^2 x \frac{\partial^2 }{\partial x^2} \frac{\partial^2  x\xi}{\partial \theta_h^2}, \mbox{   or   } \\
 &=& - \frac{1}{a^2\tau_A^2} x \frac{\partial^2 x\xi}{\partial x^2}, \label{HK} \mbox{   when } \label{xi eq} \\
 \xi&=&\bar{\xi} \cos\theta_h \mbox{  and} \\
 \tau_A &\equiv& \frac{R_0}{M\iota'r_s V_A}
\end{eqnarray}
is the characteristic time for the propagation of a shear Alfv\'en wave, typically of order a micro-second in a large tokamak.

Hahm and Kruskal gave the solution for the displacement, Equation (\ref{xi eq}), using the sine function, Si$(\zeta)$;
\begin{eqnarray}
\bar{\xi}&=&\frac{2\xi_d}{\pi} \mbox{Si}\left( \frac{x}{\Delta}\right),  \mbox{   where   } \\
\Delta(t) &\equiv& r_s\frac{\tau_A}{t};  \label{current width} \\
\mbox{Si}(\zeta) &\equiv& \int_0^\zeta \frac{\sin(u)}{u} du;\\
\mbox{Si}(\infty) &=& \frac{\pi}{2};  \\
\mbox{Si}(\zeta) &=& \zeta - \frac{\zeta^3}{18} + \frac{\zeta^5}{600} + \cdots; \\\\
\mbox{Si}(\zeta) &\leq& \mbox{Si}(\pi) =1.851937\cdots. 
 \end{eqnarray}
 The properties of the sine function, Si$(\zeta)$, can be found on a number of web sites.

The distance $\Delta(t)$ is the distance scale of the shielding current at the time $t$ after the perturbation is initiated. 

Let $\tilde{A}$ be the part of $A$ that depends on time and $\theta_h$, then in the Hahm-Kruskal approximation
\begin{eqnarray}
 \tilde{A} &=& - \frac{\mu_0\bar{J}_0}{2} x \bar{\xi}\\
 &=& - \xi_d \frac{\mu_0\bar{J}_0}{\pi} \frac{x^2t}{\Delta(t)} \frac{\mbox{Si}(\zeta)}{\zeta}.
 \end{eqnarray}
 
 When $x<\Delta(t)$, which means within the region in which a shielding current arises, $\zeta=x/\Delta<1$ and Si$(\zeta)/\zeta\approx 1$.  Consequently,
 \begin{eqnarray}
\bar{\xi} &\approx& \frac{2\xi_d}{\pi \Delta(t)}x; \\
  \frac{\tilde{A}}{A_0} &\approx& - \frac{4}{\pi}\frac{\xi_d}{\Delta(t)}; \\
 \tilde{j}_z &\equiv& -\frac{1}{\mu_0}\frac{\partial^2\tilde{A}}{\partial x^2}\\ \nonumber\\
 &\approx& \frac{2}{\pi}\bar{J}_0 \frac{\xi_d}{\Delta(t)}. \label{j-density}
 \end{eqnarray}
 Within the spatial region of validity of the Hahm-Kulsrud approximation, the shielding current is a spatial constant when $x<<\Delta(t)$. Their solution is invalid unless $\big|\bar{\xi}/2x\big|=\xi_d/(\pi\Delta)<<1$, which means when $\Delta(t)\lesssim\xi_d$.  The displacement $\xi_d$ is defined by the asymptotic value of $\bar{\xi}$ when $x>>\Delta(t)$, which means outside of the region in which shielding currents flow.  The displacement is independent of $x$ in the region $r_s>>x>>\Delta(t)$ and equal to $\xi_d$, where $r_s$ is the radius of the rational surface.
 
 


\begin{thebibliography}{99}

\bibitem{ITER-shutdown2018} P. C. de Vries, T. C. Luce, Y. S. Bae, S. Gerhardt, X. Gong, Y. Gribov, D. Humphreys, A. Kavin, R. R. Khayrutdinov, C. Kessel, S. H. Kim, A. Loarte, V.E. Lukash, E. de la Luna, I. Nunes, F. Poli, J. Qian, M. Reinke, O. Sauter, A. C. C. Sips, J.  A. Snipes, J. Stober, W. Treutterer, A. A. Teplukhina, I. Voitsekhovitch, M. H. Woo, S. Wolfe, L. Zabeo, the Alcator C-MOD team, the ASDEX Upgrade team, the DIII-D team, the EAST team, JET contributors, the KSTAR team, the NSTX-U team and the TCV team and ITPA IOS members and experts, \emph{Multi-machine analysis of termination scenarios with comparison to simulations of controlled shutdown of ITER discharges}, Nucl. Fusion \textbf{58}, 026019 (2018).


\bibitem{Boozer:steering} A. H. Boozer, \emph{Plasma steering to avoid disruptions in ITER and tokamak power plants}, Nucl. Fusion \textbf{61}, 054004 (2021).

 \bibitem{Eidietis:2021} N. W. Eidietis, \emph{Prospects for Disruption Handling in a Tokamak-based Fusion Reactor}, Fusion Science and Technology, \textbf{77}, 738 (2021).

\bibitem{Parker-Krook:1956} E. N. Parker and M. Krook, \emph{Diffusion and severing of magnetic lines of force}, Ap. J. \textbf{124}, 214 (1956).

\bibitem{de Vries:2016} P.C. de Vries, G. Pautasso, E. Nardon, P. Cahyna, S. Gerasimov, J. Havlicek, T.C. Hender, G.T.A. Huijsmans, M. Lehnen, M. Maraschek, T. Markovi\v{c}, J.A. Snipes and the COMPASS Team, the ASDEX Upgrade Team and JET Contributors, \emph{Scaling of the MHD perturbation amplitude required to trigger a disruption and predictions for ITER},  Nucl. Fusion \textbf{56}, 026007 (2016).

\bibitem{L/R} A. H. Boozer, \emph{Pivotal issues on relativistic electrons in ITER}, Nucl. Fusion \textbf{58}, 036006 (2018).


\bibitem{NIMROD} C.R. Sovinec, A.H. Glasser, D.C. Barnes, T.A. Gianakon, R.A. Nebel, S.E. Kruger, D.D. Schnack, S.J. Plimpton, A. Tarditi, M.S. Chu and the NIMROD Team, \emph{Nonlinear magnetohydrodynamics simulation using high-order finite elements}, J. Comp. Phys., \textbf{195}, 355 (2004).

\bibitem{M3D-C1} N. M. Ferraro, B. C. Lyons, C. C. Kim, Y. Q. Liu, and S. C. Jardin, \emph{3D two-temperature magnetohydrodynamic modeling of fast thermal quenches due to injected impurities in tokamaks}, Nucl. Fusion \textbf{59} 016001 (2019).

\bibitem{JOREK} M. Hoelzl, G.T.A. Huijsmans, S.J.P. Pamela, M. B\'ecoulet, E. Nardon, F.J. Artola, B. Nkonga, C.V. Atanasiu, V. Bandaru, A. Bhole, D. Bonfiglio , A. Cathey, O. Czarny, A. Dvornova, T. Feh\'er, A. Fil, E. Franck, S. Futatani, M. Gruca, H. Guillard, J.W. Haverkort, I. Holod, D. Hu, S.K. Kim, S.Q. Korving, L. Kos, I. Krebs, L. Kripner, G. Latu, F. Liu, P. Merkel, D. Meshcheriakov, V. Mitterauer, S. Mochalskyy, J.A. Morales, R. Nies, N. Nikulsin, F. Orain, J. Pratt, R. Ramasamy, P. Ramet, C. Reux, K. S\"arkim\"aki, N. Schwarz, P. Singh Verma, S.F. Smith, C. Sommariva, E. Strumberger, D.C. van Vugt,
M. Verbeek, E. Westerhof, F. Wieschollek, and J. Zielinski, \emph{The JOREK non-linear extended MHD code and applications to large-scale instabilities and their control in magnetically confined fusion plasmas}, Nucl. Fusion \textbf{61}, 065001 (2021).

\bibitem{Schindler:1988} K. Schindler, M. Hesse, and J. Birn, \emph{General magnetic reconnection, parallel electric-fields, and helicity}, Journal of Geophysical Research---Space Physics \textbf{93}, 5547 (1988).

\bibitem{plasmoid} D. A. Uzdensky, N. F. Loureiro, and A. A. Schekochihin, \emph{Fast Magnetic Reconnection in the Plasmoid-Dominated Regime}, Phys. Rev. Lett. \textbf{105}, 235002 (2010).

\bibitem{Boozer:JPP} A. H. Boozer, \emph{Fast Magnetic Reconnection}, J. Plasma Phys. Colloquium Series, 9 September 2021, $\big<$\url{https://mediacentral.princeton.edu/id/1_qgbjho9t}$\big>$.

\bibitem{RDR-kink} M. N. Rosenbluth, R. Y. Dagazian, and P. H. Rutherford, \emph{Nonlinear properties of the internal m=1  kink instability in the cylindrical tokamak}, Phys. Fluids \textbf{16}, 1894 (1973).

\bibitem{Hahm-Kulsrud} T. S. Hahm and R. M. Kulsrud, \emph{Forced magnetic reconnection}, Phys. Fluids \textbf{28}, 2412 (1985).

\bibitem{Boozer-Pomphrey} A.  H. Boozer, and N. Pomphrey, \emph{Current density and plasma displacement near perturbed rational surfaces}, Phys. Plasmas \textbf{17}, 110707 (2010).

\bibitem{Zhou:2016} Y. Zhou, Y-M Huang, H. Qin, and A. Bhattacharjee, \emph{Formation of current singularity in a topologically constrained plasma}, Phys. Rev. E \textbf{93}, 023205 (2016).

\bibitem{Zhou:2019} Y. Zhou, Y-M Huang, A. H. Reiman, H. Qin, and A. Bhattacharjee, \emph{Magnetohydrodynamical equilibria with current singularities and continuous rotational transform}, Phys. Plasmas \textbf{26}, 022103 (2019).

\bibitem{Huang:2021} Y.-M. Huang, S. R. Hudson, J. Loizu, Y. Zhou, and A. Bhattacharjee, \emph{Numerical approach to $\delta$-function current sheets arising from resonant magnetic perturbations}, arXiv (August 2021),  $\big<$\url{https://arxiv.org/pdf/2108.09327.pdf}$\big>$.


\bibitem{Boozer:RMP} A. H. Boozer, \emph{Physics of magnetically confined plasmas}, Rev. Mod. Phys. \textbf{76}, 1071 (2004).


\bibitem{Rec-example} A. H. Boozer and T. Elder, \emph{Example of exponentially enhanced magnetic reconnection driven by a spatially bounded and laminar ideal flow}, Phys. Plasmas \textbf{28}, 062303 (2021).

\bibitem{Boozer:rec-phys} A. H. Boozer, \emph{Magnetic reconnection and thermal equilibration}, Phys. Plasmas \textbf{28}, 032102 (2021).

\bibitem{Amitava:2004} A. Bhattacharjee, \emph{Impulsive magnetic reconnection in the earth's magnetotail and the solar corona}, Annu. Rev. Astron. Astrophys. \textbf{42}, 365 (2004).

\bibitem{Parker:1973} E. N. Parker, \emph{The reconnection rate of magnetic fields}, Ap. J. \textbf{180}, 247 (1973).

\bibitem{Pariat-Antiochos} E. Pariat, S. K. Antiochos, and C. R. DeVore, \emph{A model for solar polar jets}, Ap. J. \textbf{691}, 61 (2009).

\bibitem{Boozer:null-X} A. H. Boozer, \emph{Magnetic Reconnection with null and X-points}, Phys. Plasmas \textbf{26}, 122902 (2019).

\bibitem{Photospheric flows:2018} S.-H. Park, J. A. Guerra, P. T. Gallagher, M. K. Georgoulis, and D. S. Bloomfield, \emph{Photospheric Shear Flows in Solar Active Regions and Their Relation to Flare Occurrence}, Solar Phys \textbf{293}, 114 (2018).

\bibitem{Boozer:acc} A. H. Boozer, \emph{Particle acceleration and fast magnetic reconnection}, Phys. Plasmas  \textbf{26}, 082112 (2019).

\bibitem{Boozer:j-||} A. H. Boozer, \emph{Flattening of the tokamak current profile by a fast magnetic reconnection with implications for the solar corona}, Phys. Plasmas \textbf{27}, 102305 (2020).

\bibitem{Paz-Soldan:2020} C. Paz-Soldan, P. Aleynikov, E.M. Hollmann, A. Lvovskiy, I. Bykov, X. Du, N.W. Eidietis, and D. Shiraki, \emph{Runaway electron seed formation at reactor-relevant temperature}, Nucl. Fusion \textbf{60}, 056020 (2020).

\bibitem{Boozer:separation} A. H. Boozer, \emph{Separation of magnetic field lines}, Phys. Plasmas \textbf{19}, 112901 (2012).


\bibitem{Lazarian:1999} A. Lazarian and E. T. Vishniac, \emph{Reconnection in a weakly stochastic field}, Ap. J. \textbf{517}, 700 (1999).

\bibitem{Eyink:2011} G. L. Eyink, A. Lazarian, E. T. Vishniac, \emph{Fast magnetic reconnection and spontaneous stochasticity}, Ap. J. \textbf{743}, 51 (2011).

\bibitem{Eyink:2015} G. L. Eyink, \emph{Turbulent general magnetic reconnection}, Ap. J \textbf{807} 137 (2015).
 
\bibitem{Matthaeus:2015} W. H. Matthaeus, M. Wan, S. Servidio, A. Greco, K. T. Osman, S. Oughton, and P. Dmitruk, \emph{Intermittency, nonlinear dynamics and dissipation in the solar wind and astrophysical plasmas}, Phil. Trans. R. Soc. A \textbf{373} 20140154 (2015).

\bibitem{Matthaeus:2020} S. Adhikari, M. A. Shay, T. N. Parashar, P. S. Pyakurel, W. H. Matthaeus, D. Godzieba, J. E. Stawarz, J. P. Eastwood, and J. T. Dahlin,  \emph{Reconnection from a turbulence perspective}, Phys. Plasmas \textbf{27}, 042305  (2020).

\bibitem{Lazarian:2020rev} A. Lazarian, G. L. Eyink, A. Jafari, G. Kowal, H. Li, S-Y Xu, and E. T. Vishniac, \emph{3D turbulent reconnection: Theory, tests, and astrophysical implications}, Phys. Plasmas \textbf{27}, 012305 (2020).

\bibitem{Priest:2016} E. Priest, \emph{MHD structures in three-dimensional reconnection}, volume \textbf{427}, page 101, \emph{Book Series Astrophysics and Space Science Library},  \emph{Magnetic reconnection: concepts and applications}, (Springer International Publishing 2016, edited by WalterGonzalez and Eugene Parker)

\bibitem{Reid:2020} J. Reid, C. E. Parnell, A. W. Hood, and P. K. Browning, \emph{Determining whether the squashing factor, Q, would be a good indicator of reconnection in a resistive MHD experiment devoid of null points}, Astronomy and Astrophysics \textbf{633}, A92 (2020).

\bibitem{Boozer:coordinates} A. H. Boozer, \emph{Plasma equilibrium with rational magnetic surfaces}, Phys. Fluids \textbf{24}, 1999 (1981).

\bibitem{Grad:1967} H. Grad, \emph{Toroidal containment of a plasma}, Phys. Fluids \textbf{10}, 137 (1967).

 \bibitem{Newcomb} W. A. Newcomb, \emph{Motion of magnetic lines of force}, Ann. Phys. \textbf{3}, 347 (1958).


\bibitem{Lighthill:1986} J. Lighthill, \emph{The recently recognized failure of predictability in Newtonian dynamics}, Proceedings of the Royal Society, Series A, \textbf{407}, 35 (1986).

\bibitem{Aref:1984} H. Aref, \emph{Stirring by chaotic advection}, Journal of Fluid Mechanics \textbf{143}, 1 (1984).

\bibitem{Aref:2017} H. Aref, J. R. Blake, Marko Budisi\'c, S. S. S. Cardoso, J. H. E. Cartwright, H. J. H. Clercx, K. El Omari, U. Feudel, R. Golestanian, E. Gouillart, G. J. F. van Heijst, T. S. Krasnopolskaya, Y. Le Guer, R. S. MacKay, V. V. Meleshko, G. Metcalfe, I. Mezi\'c, A. P. S. de Moura, O. Piro, M. F. M. Speetjens, R. Sturman, J.-L. Thiffeault, and I. Tuval, \emph{Frontiers of chaotic advection}, Rev. Mod. Phys. \textbf{89}, 025007 (2017).

\bibitem{Courant-Snyder} E. D. Courant and H. S. Snyder, \emph{Theory of the alternating-gradient synchrotron}, Ann. Phys. \textbf{3}, 1 (1958), Appendix B.

\bibitem{Boozer:space} A. H. Boozer, \emph{Magnetic reconnection in space}, Phys. Plasmas \textbf{19}, 092902 (2012).

\bibitem{Stern:1966} D. P. Stern, \emph{Motion of magnetic field lines}, Space Sci. Rev. \textbf{6}, 147 (1966).

\bibitem{Zhou:2021} Y. Zhou, private communication, April 28, 2021.



\end{thebibliography}
\end{document}